\documentclass[aps,prd,twocolumn,eqsecnum,amsmath,nofootinbib,preprintnumbers]{revtex4}%
\usepackage{graphicx}
\usepackage{xcolor}
\usepackage{float}
\definecolor{wred}{rgb}{0,0.618,0.0}
\definecolor{wblue}{rgb}{.0,0.0,0.618}
\definecolor{wgreen}{rgb}{.0,0.618,0.0}
\usepackage{tikz}
\usepackage{multirow}
\usetikzlibrary{decorations.pathmorphing,calc}
\usepackage[
colorlinks=true,
filecolor=black,
linkcolor=wblue,
citecolor=wgreen,
pdfstartview=FitH,
pdfpagemode=None,
bookmarksopen=true,
urlcolor=blue
]{hyperref}

\begin{document}
	\preprint{\hfill {\small {1}}}
	
	\title{Time Dependence of Holographic Complexity in Gauss-Bonnet Gravity }
	\author{Yu-Sen An$^{a,b}$\footnote{Email: anyusen@itp.ac.cn}, Rong-Gen Cai$^{a,b}$\footnote{Email: cairg@itp.ac.cn} and Yuxuan Peng$^{a}$\footnote{Email:  yxpeng@itp.ac.cn}}

	\affiliation{$^a$
		CAS Key Laboratory of Theoretical Physics, Institute of Theoretical Physics,
		Chinese Academy of Sciences, Beijing 100190, China}
	\affiliation{$^b$ 
		School of Physical Sciences, University of Chinese Academy of Sciences,
		Beijing 100049, China}
	\begin{abstract}
		We study the effect of the Gauss-Bonnet term on the  complexity growth rate  of  dual field theory using the ``Complexity--Volume'' (CV) and CV2.0 conjectures. We investigate the late time value and full time evolution of the complexity growth rate of the  Gauss-Bonnet black holes with horizons with zero curvature ($k=0$), positive curvature ($k=1$)  and negative curvature ($k=-1$) respectively. 
		For the $k=0$ and $k=1$ cases we find that the Gauss-Bonnet term suppresses the growth rate as expected, while in the $k=-1$ case the effect of the Gauss-Bonnet term may be opposite to what is expected. The reason for it is briefly discussed, and the comparison of our results to the result obtained by using the ``Complexity--Action'' (CA) conjecture is also presented.
		{We also briefly investigate two proposals applying some generalized volume functionals dual to the complexity in higher curvature gravity theories, and find their behaviors are different for $k=0$ at late times.}
	\end{abstract}
	\maketitle
	\tableofcontents
	\section{Introduction}
	The Anti-de Sitter/Conformal Field Theory (AdS/CFT) correspondence relates a gravity theory in an asymptotically AdS spacetime, often referred to the bulk, to a conformal field theory without gravity living on the boundary of this spacetime \cite{Maldacena:1997re, Gubser:1998bc, Witten:1998qj, Aharony:1999ti}.
	This correspondence is the most important realization of the holographic principle \cite{tHooft:1993dmi,Susskind:1994vu}.
	Studies on the relation between gravity and quantum information in the context of AdS/CFT correspondence has been an important topic in recent years, and
	one  famous subject in this direction is the holographic entanglement entropy proposed by Ryu and Takayanagi \cite{Ryu:2006bv}. 
	The Ryu-Takayanagi formula allows one to express the entanglement entropy of a conformal field theory in some subregion of the boundary by the minimal area of a bulk co-dimension two surface anchored at the boundaries of the subregion.
	Interestingly, the holographic duality also suggests that the dynamics of the bulk spacetime emerges from quantum entanglement on the boundary \cite{Lashkari:2013koa,Faulkner:2013ica,Swingle:2014uza}.
	
	More recently the quantum complexity, another quantity in quantum information theory, attracted a lot of attention.
	The quantum complexity of a certain state describes how many simple operations (quantum gates) at least are needed to obtain this state from some chosen reference state.
	For a discrete system composed of a number of quantum bits, the complexity measures its ability of computation.
	The concept of quantum complexity in quantum field theory is still unclear. 
	However, there are two potential holographic descriptions for it in the bulk spacetime.
	The first one is that the volume of the Einstein-Rosen bridge of an eternal AdS black hole with two asymptotic boundaries is proportional to the quantum complexity of the dual field theory.
	This conjecture is often called the ``Complexity--Volume'' (henceforth CV) conjecture, originally proposed by Susskind in \cite{Stanford:2014jda} in the purpose of finding the boundary dual of the size of the Einstein-Rosen bridge which increases for an exponentially long time.
	The mathematical relation is 
	\begin{eqnarray}\label{CV}
		C_V=\frac{{\rm max}\left[V\right]}{G\ell}\,.
	\end{eqnarray}
	Strictly speaking, $C_V$ denotes the complexity of a specific state  of a system composed of two identical copies of conformal field theories (CFTs).
	The two identical copies are denoted by the ``left'' part $\rm CFT_{L}$ and the ``right'' part $\rm CFT_{\rm R}$ respectively, and the specific state $|\Psi\rangle$ is the thermofield double (TFD) state
	\begin{eqnarray}
		|TFD\rangle \equiv \frac{1}{\sqrt{Z(\beta)}}\sum_{n}e^{-\frac{\beta E_n}{2}}|n\rangle_{\rm L}|n\rangle_{\rm R}
	\end{eqnarray}
	evolved to some certain time denoted by $\tau_{\rm L}$ and $\tau_{\rm R}$, the time coordinates of the two copies:
	\begin{eqnarray}
		|\Psi\rangle = e^{-i(H_{\rm L} \tau_{\rm L} + H_{\rm R} \tau_{\rm R})}|TFD\rangle\,,
	\end{eqnarray}
	where $H_{\rm L}$ and $H_{\rm R}$ are the Hamiltonian operators of the two subsystems.
	The symmetry of this state allows us to choose $\tau_{\rm L} = \tau_{\rm R}$.
	The states $|n\rangle_{\rm L}$ and $|n\rangle_{\rm R}$ are the energy eigenstates of energy $E_n$ of the two copies and the sum is over all the energy eigenstates. 
	When the degrees of freedom of one copy is traced out, the reduced density matrix of the other copy is just that of a thermal state with temperature $1/\beta$, and $Z(\beta)$ is the corresponding partition function.
	The holographic dual of the state $|\Psi\rangle$ is an eternal AdS black hole \cite{Maldacena:2001kr} whose Penrose diagram is shown in Figure \ref{CVPenrose}.
	The $V$ in the right hand side of Eq.(\ref{CV}) is the volume of a certain co-dimension one surface connecting the constant time slices $\tau = \tau_{\rm L}$ and $\tau = \tau_{\rm R}$ on both boundaries, and the symbol ``max'' means that we choose the surface such that the volume is maximal\footnote{Another interpretation of the volume other than the complexity was proposed by Ref.\cite{MIyaji:2015mia}.}.
	The constant $G$ is the Newton constant, and $\ell$ is some length scale, usually chosen to be the curvature radius of the AdS spacetime. 
	
	The other conjecture of holographic complexity is the so-called ``Complexity--Action'' (CA) conjecture proposed in \cite{Brown:2015bva,Brown:2015lvg}. It relates the complexity of the boundary CFT state
	to the gravitational action in the so-called Wheeler-DeWitt (henceforth WdW) patch in the bulk, and successfully avoids the unclear dimensionful parameter $\ell$ appearing in the CV conjecture.
	As is shown in Figure \ref{CAPenrose}, the WdW patch is the  domain of dependence of any spacelike hypersurface connecting the two time slices on $\tau_{\rm L}$ and $\tau_{\rm R}$ on the boundaries. The precise relation between the complexity $C_A$ and the action $I$ is
	\begin{eqnarray}\label{CA}
		C_A=\frac{I}{\pi \hbar}\,,
	\end{eqnarray}
	where $\hbar$ is just the reduced Planck constant.
	These studies motivated  a lot of discussions on holographic complexity \cite{Alishahiha:2015rta,Ben-Ami:2016qex,Chapman:2016hwi,Carmi:2016wjl,Kim:2017lrw,Gan:2017qkz}.
	
	There has also been an upsurge in the study of complexity on the field theory side recently \cite{Jefferson:2017sdb,Chapman:2017rqy,Yang:2017nfn,Khan:2018rzm,Yang:2018nda,Hackl:2018ptj,Sinamuli:2018jhm,Jiang:2018gft}, just after the holographic conjecture was proposed.
	
	One important property of the complexity of a quantum system is the bound on its growth rate proposed in \cite{Brown:2015lvg} 
	\begin{eqnarray}\label{Lloyd}
		\frac{\mathrm{d}C}{\mathrm{d}\tau}\leq \frac{2 E}{\pi \hbar}\,,
	\end{eqnarray}
	which is often referred to as the ``Lloyd bound'' since it was inspired by the bound conjectured by Lloyd \cite{Lloyd}.
	Here $E$ is the total energy of the system.
	Verification of this bound on the gravity side using the CA conjecture was first done in \cite{Brown:2015bva,Brown:2015lvg}.
	The results therein showed that at late times the complexity growth rate approaches a certain limit, and for the neutral AdS black holes in Einstein gravity theory the late time limit is just the right hand side of Eq.(\ref{Lloyd}). The bound for rotating and charged black holes is tighter than Eq.(\ref{Lloyd}).
	Afterwards Cai  {\it et al.}\cite{Cai:2016xho} presented a universal formula for the action growth expressed in terms of some thermodynamical quantities associated with the outer and inner horizons of the AdS black holes. For related discussions, see also the papers \cite{Yang:2016awy,Huang:2016fks,Cai:2017sjv,Ge:2017rak} and references therein.
	There have also been a lot of discussions on the complexity growth rate in different spacetime settings and different gravity theories other than Einstein gravity \cite{Pan:2016ecg,Alishahiha:2017hwg,Wang:2017uiw,Guo:2017rul,Miao:2017quj,Ghodrati:2017roz,Qaemmaqami:2017lzs,Sebastiani:2017rxr,Moosa:2017yvt,Moosa:2017yiz,Chapman:2018dem,Auzzi:2018zdu,Du:2018uua,Momeni:2016ira,Jiang:2018pfk}.
	
	While the CA conjecture passed several non-trivial tests, there are also some problems suggesting the correction of this conjecture, such as the violation of the Lloyd bound in full time evolution of the action found in the paper \cite{Carmi:2017jqz}.
	The results therein show that the growth rate in the CV conjecture grows monotonically with time, and it approaches a certain bound from below\footnote{If we appropriately choose the constant $\ell$, then this bound is the Lloyd bound \cite{Kim:2017qrq}, as will also be mentioned in this paper.}, while in the CA conjecture the complexity growth rate grows with time and exceeds the Lloyd bound, and then approaches the bound from above at late times. 
	Recently several studies also found some violation of this bound in the late time limit in the CA conjecture, such as \cite{An:2018xhv,Swingle:2017zcd,Alishahiha:2018tep}. 

	Besides the usual CV and CA duality, there are also various interesting modified holographic proposals discussed in the papers \cite{Bueno:2016gnv,Alishahiha:2017hwg,Abt:2017pmf,Fan:2018wnv,Couch:2016exn}. In Ref.\cite{Couch:2016exn}, based on the black hole chemistry, the authors proposed that the complexity should be dual to the spacetime volume of the WdW patch	
	\begin{eqnarray}
		C_V=\frac{1}{\hbar} P\times V_{\rm WdW}\,,
	\end{eqnarray}
which is called the ``CV2.0'' conjecture. In Ref.\cite{Bueno:2016gnv,Alishahiha:2017hwg}, the authors proposed that the volume should be modified in the presence of the higher curvature terms. In Ref.\cite {Alishahiha:2017hwg}, the authors discussed two possible forms of the modified volume in critical gravity. It should be noted that these proposals need to be investigated in other higher curvature gravity theories to test whether they are true . 
		
Gauss-Bonnet gravity is an important generalization of Einstein gravity when we go to higher curvature case. 
It is a special case of the Lovelock gravity theory \cite{Lovelock:1971yv}, the general second-order covariant gravity theory in dimensions higher than four. Moreover, the Gauss-Bonnet term can be regarded as corrections from the heterotic string theory \cite{Gross:1986iv,Zumino:1985dp}. 

  It is an important task to test the various holographic complexity conjectures beyond Einstein gravity, and the Einstein-Gauss-Bonnet black hole is a natural testing ground. A straightforward way of doing this is to generalize the work \cite{Carmi:2017jqz} to Einsein-Gauss-Bonnet gravity.
	In work \cite{Carmi:2017jqz} the full time dependence of holographic complexity growth rate for Einstein gravity is calculated for both CA and CV conjectures.
	However, when generalizing to the higher curvature case one will need the proper boundary and joint terms of the gravitational action action.
	{For the Einstein gravity these terms were first introduced in the paper \cite{Lehner:2016vdi}.
	Ref.\cite{Cano:2018ckq} studied action terms at joints between certain boundary segments of a WdW patch in Lovelock gravity,
	and by using the results therein the late time complexity growth rate of the black holes in Lovelock gravity was obtained in \cite{Cano:2018aqi}.  
	Ref. \cite{Chakraborty:2018dvi} gave the action terms on the null boundaries for Lovelock gravity, and these terms may be used to derive the full-time behavior of the complexity in Lovelock gravity.}

	Despite several works concerning the growth rate of complexity in Gauss-Bonnet case using the CA proposal, the analysis using various volume proposals for complexity is still lacked. Moreover, as argued in \cite{Brown:2015lvg}, since such corrections weaken the interaction of the boundary field theory, the complexity growth rate will be reduced. This paper will examine the growth rate of holographic complexity of the AdS black holes in the Einstein-Gauss-Bonnet (henceforth EGB) gravity theory using various volume proposals.  By comparing various volume proposals to the action proposal, we try to judge which proposal is better according to their growth behavior and the picture given in \cite{Brown:2015lvg} .



	
	This paper is organized as follows. 
	In Section \ref{GBBH}, we give a brief introduction of the Einstein-Gauss-Bonnet black hole, which is the spacetime where our computation is carried.
	In Section \ref{Calc}, we investigate the complexity growth rate using the CV proposal. We plot the late time and full time dependence of the complexity growth rate for black holes with different horizon curvatures. The charged case is also discussed. 
	In Section \ref{CV2}, we briefly investigate the complexity growth using the ``CV2.0 conjecture'' first proposed in \cite{Couch:2016exn}. 
	In Section \ref{GenVol}, we investigate two proposals of generalized volumes dual to the complexity \cite{Bueno:2016gnv,Alishahiha:2017hwg} in the presence of higher curvature corrections, and plot the late time complexity growth rate in $k=0$ for different coupling $\alpha$s, and the two proposals shows very different properties.
	In Section \ref{Discuss}, we summarize our results and give some conclusions and discussions.

	\section{ AdS Black Hole Solutions in Einstein-Gauss-Bonnet Gravity Theory}\label{GBBH}
	The action of EGB gravity theory with a negative cosmological constant and a Maxwell field is
	\begin{eqnarray}
		S &= &\frac{1}{16 \pi G}\int \mathrm{d}^{d+1} x \sqrt{-g}\left[ R + \frac{d(d-1)}{L^2}\right.\nonumber\\
		&&\left. + \alpha (R^2 -4 R_{ab}R^{ab} + R_{abcd}R^{abcd}) \right]\nonumber\\
		&&-\frac{1}{4\pi}\int \mathrm{d}^{d+1} x \sqrt{-g}F_{ab}F^{ab}
		\,.
	\end{eqnarray}
	The spacetime dimension $D=d+1\ge 5$ since in four dimensions the Gauss-Bonnet  term is just a topological term. $G$ is the Newton constant and the relation between the cosmological constant $\Lambda$ and the curvature length scale $L$ is $\Lambda = -d(d-1)/(2L^2)$. 
	The Gauss-Bonnet coupling is denoted by $\alpha$ and to avoid verboseness we use the symbol $\tilde{\alpha} \equiv \alpha (d-2)(d-3)$.
	In the heterotic string theory $\alpha$ is positive \cite{Boulware:1985wk}, while for $d=4$ the causality of the boundary CFT gives the constraint $-7/36\leq\tilde{\alpha}/L^2\leq 9/100$ \cite{Brigante:2007nu,Brigante:2008gz,Buchel:2009tt,Hofman:2009ug}. 
	Considering these facts, we will restrict ourselves to the parameter range\footnote{Though the causality constraint is obtained only for the $k=0$ black branes, we also consider this constraint for $k=\pm 1$ cases.} $0\leq\tilde{\alpha}/L^2\leq 9/100$ when $d=4$.
	The electromagnetic field strength appears as the tensor $F_{ab}$ as usual, while in most part of this paper we only study the case of neutral black holes without electromagnetic fields.
	The $(d+1)$-dimensional static black hole solutions of the above action are described by the line element
	\begin{equation}\label{staticbh}
		\mathrm{d}s^2 = -f(r)\mathrm{d}\tau^2 + \frac{1}{f(r)}\mathrm{d}r^2 + r^2 h_{ij}\mathrm{d}x^i \mathrm{d}x^j
	\end{equation}
	where\footnote{Here we largely follow the notations in Ref. \cite{Torii:2005nh}.}
	\begin{eqnarray}\label{blakening}
		f(r) = k+\frac{r^2}{2 \tilde{\alpha}} \left(1 \mp \sqrt{1+4\tilde{\alpha}\left(\frac{\tilde{M}}{  r^d}-\frac{1}{L^2} - \frac{  \tilde{Q}^2}{r^{2d-2}}\right)}\right)\,,\nonumber\\
	\end{eqnarray}
	and
	\begin{eqnarray}
		\tilde M \equiv \frac{16 \pi   G M}{(d-1) \Omega_{k,d-1} }\,,\qquad \tilde{Q}^2  \equiv  \frac{2 (d-2) G Q^2}{(d-1)}\,.
	\end{eqnarray}
	In the expressions above the black hole mass and charge are $M$ and $Q$ respectively, and $\Omega_{k,d-1}$ is the volume of the maximally symmetric $(d-1)$-dimensional submanifold with the metric $h_{ij}$, and this submanifold can have zero, positive or negative curvature, corresponding to the parameter $k$ equal to $1$, $0$ or $-1$ respectively. The solution with $k=1$ was found in the paper \cite{Boulware:1985wk}, and the solutions with $k=0, -1$ were first given in the paper \cite{Cai:2001dz} and the charged solutions were found in \cite{Wiltshire:1985us}.
	For these three cases the horizons can have spherical, planar and hyperbolic topology respectively.
	There are two branches of solutions due to the choice of the sign in front of the square root in the expression (\ref{blakening}), and throughout this paper we study the branch with a minus sign ``$-$'' which approaches the asymptotically AdS black hole solution in Einstein gravity theory in the limit $\alpha\rightarrow 0$.
	The mass parameter can be expressed in terms of the horizon radius $r_h$ and the other parameters as
	\begin{eqnarray}\label{M}
		M = \frac{(d-1)\Omega_{k,d-1} r_h^{d-2}}{16\pi G}\left( k+\frac{\tilde{\alpha}k^2}{r_h^2} + \frac{r_h^2}{L^2} + \frac{2(d-2)G{Q}^2}{(d-1)r_h^{2d-4}} \right)\,.\nonumber
	\end{eqnarray}
	The temperature $T$ and the entropy $S$ of the black hole are \cite{Cai:2013qga}
	\begin{eqnarray}
		&&T=\frac{d r_h^4+(d-2) k  L^2 r_h^2 + {\tilde{\alpha}} (d-4) k ^2 L^2}{4 \pi  L^2 r_h \left(2 \tilde{\alpha} k +r_h^2\right)}\nonumber\\
		&&\qquad-\frac{(d-2)^2 G Q^2 r_h^{5-2 d}}{2 \pi  (d-1) \left(2 \tilde{\alpha}k +r_h^2\right)}\,,\nonumber\\
		&&S= \frac{\Omega_{k,d-1} r_h^{d-1}}{4G}\left( 1 + \frac{d-1}{d-3} \frac{2 \tilde{\alpha}k}{r_h^2} \right)\,.
	\end{eqnarray}

	\section{The Growth Rate of Complexity in the CV Conjecture}\label{Calc}
	\subsection{The Method}\label{Method}
\begin{figure}
				\includegraphics[scale=0.7]{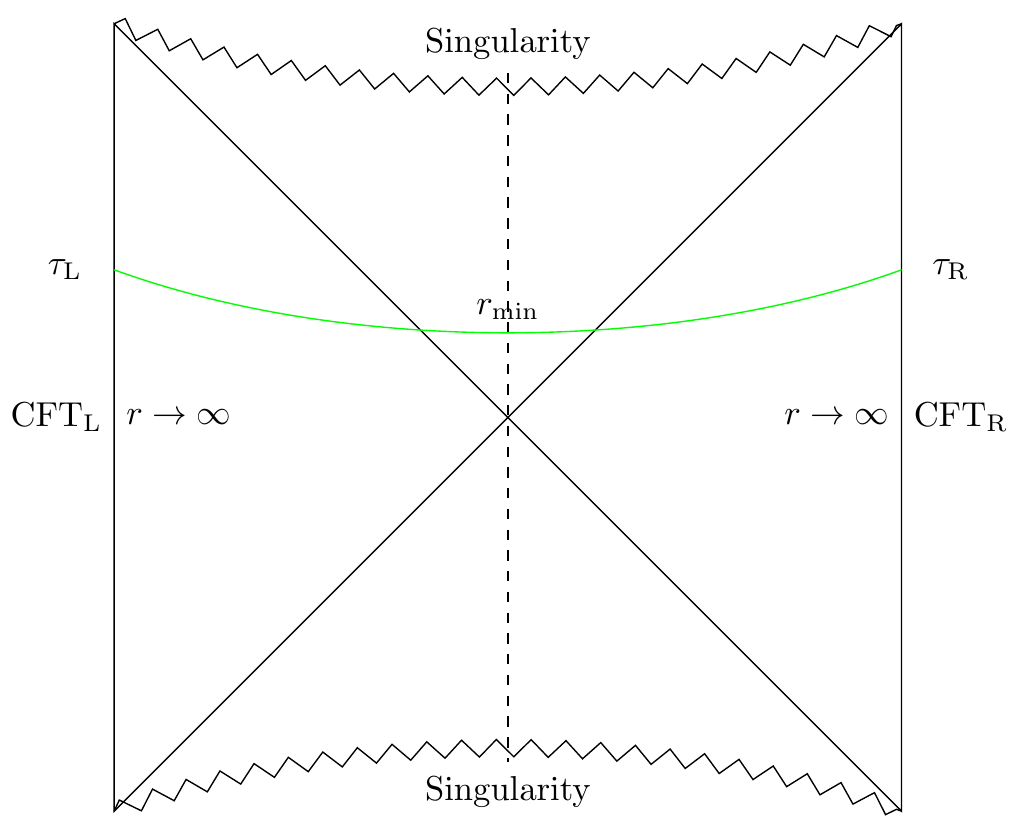}
						\caption{The Penrose diagram of an eternal neutral AdS black hole. The co-dimension one spacelike surface with maximal volume connecting two boundary time slices at time $\tau_{\rm L}$ and $\tau_{\rm R}$ is shown in the diagram by green color. The two time slices on both boundaries are chosen so that the maximal surface is symmetric and $\tau_{\rm L} = \tau_{\rm R}$ without losing generality. The maximal surface can be divided into two equivalent parts at the minimal radius $r_{\rm min}$.}
						\label{CVPenrose}
		\end{figure}

	Let us first review the method of calculating the complexity growth rate for an eternal AdS black hole, following \cite{Stanford:2014jda,Carmi:2017jqz}.
	The Penrose diagram of such a spacetime is shown in Figure \ref{CVPenrose}. There are two timelike boundaries on the left and right separated by the black hole horizons, and the two copies of CFT are denoted by $\rm CFT_{\rm L}$ and $\rm CFT_{\rm R}$ respectively.
	The maximal co-dimension one surface indicated by green color connects the two spacelike slices at the boundary time $\tau_{\rm L}$ on the left side and $\tau_{\rm R}$ on the right side.
	The minimal value of the radial coordinate of the surface is denoted by $r_{\rm min}$.
	The boundary time coordinates are chosen to satisfy the relation $\tau_{\rm R} = t$ and $\tau_{\rm L} = -t$ with $t$ in (\ref{staticbh}). 
	Both $\tau_{\rm L}$ and $\tau_{\rm R}$ increase in the future direction, and one can always impose a translation along $t$ to have $\tau_{\rm L} = \tau_{\rm R}$ and the maximal surface symmetric with respect to the dashed central line ($t=0$).
	So we only need to care about the time evolution with respect to the total time $\tau=\tau_{\rm L} + \tau_{\rm R}$, and we restrict ourselves to the range $\tau\geq0$, i.e. we only consider the evolution in the upper half of the diagram.
	\begin{figure}
				\includegraphics[scale=0.7]{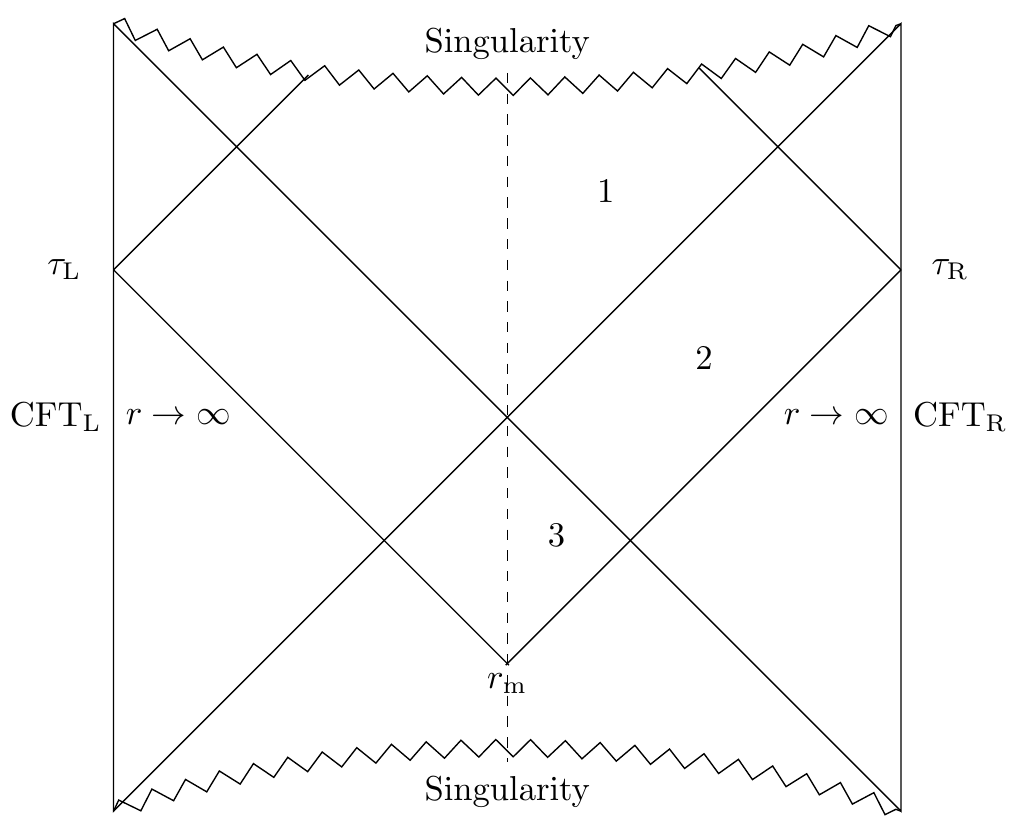}
		\caption{The Wheeler-DeWitt patch of the AdS black hole is shown in this figure. Each of the time slices at $\tau_{\rm L}$ and $\tau_{\rm R}$ acts as an endpoint of two null-sheets pointing into the bulk. The two lower sheets intersect at the point with radial coordinate $r_{\rm m}$ and the two upper ones end at the singularity. The region enclosed by the null sheets together with the boundaries and joint points is called the Wheeler-DeWitt patch.}
		\label{CAPenrose}
	\end{figure}
	To find the surface with maximal volume, the first step is to write the line element (\ref{staticbh}) in the Eddington-Finkelstein coordinates
	\begin{eqnarray}\label{Eddington}
		\mathrm{d}s^2 = -f(r)\mathrm{d}v^2 + 2\mathrm{d}v\mathrm{d}r + r^2 h_{ij}\mathrm{d}x^i \mathrm{d}x^j\,,
	\end{eqnarray}
	where
	\begin{eqnarray}
		v = t+ r^*(r)\,,\qquad \mathrm{d}r^* = \frac{\mathrm{d}r}{f(r)}\,.
	\end{eqnarray}
	Suppose the surface possesses the same maximal symmetry as the horizon does, i.e. its embedding does not depend on the coordinates $x^i$. Therefore the surface can be described by the parametric equations $v=v(\lambda)$ and $r=r(\lambda)$ with some parameter $\lambda$, then its volume is an integral in the following form
	\begin{eqnarray}
		V = 2\Omega_{k,d-1} \int_{r_{\rm min}}^{\infty} \mathrm{d} \lambda \, r^{d-1} \sqrt{-f(r) \dot{v}^2 + 2 \dot{v}\dot{r}}\,,
		\label{volume_int}
	\end{eqnarray}
	where  the dots indicate derivative with respect to $\lambda$. We have used the fact that the surface is composed of two equivalent parts and there appears the overall factor $2$.
	The integration is evaluated over the right part of the surface.
	We can define the function ${\cal L}\equiv r^{d-1} \sqrt{-f(r) \dot{v}^2 + 2 \dot{v}\dot{r}}$, and
	this function should satisfy a set of Euler-Lagrange (henceforth E-L) equations.
	Since $\cal L$ does not depend explicitly on $v$, one of the E-L equations gives that
	\begin{eqnarray}
		E=-\frac{\partial \mathcal{L}}{\partial \dot{v}} =\frac{r^{d-1} (f \dot{v}-\dot{r})}{\sqrt{-f \dot{v}^2 + 2\dot{v}\dot{r}}}
	\end{eqnarray}
	where $E$ is a constant on the whole surface (but a function of the boundary time $\tau$). We are not using the other E-L equation, but we consider the fact that the expression in Eq.(\ref{volume_int}) is reparametrization invariant.
	Thus we are free to choose $\lambda$ to keep the radial volume element fixed as follows
	\begin{equation}
		r^{d-1} \sqrt{-f\, \dot{v}^2 + 2\dot{v}\dot{r}} =1 \,.
	\end{equation}
	The two equations above simplify to
	\begin{eqnarray}
		E&=&r^{2(d-1)} \left(f(r) \dot{v}-\dot r\right), \label{tdot} \\
		r^{2(d-1)} \dot{r}^2 &=& f(r) + r^{-2(d-1)} E^2 \label{rdot},
	\end{eqnarray}
	and further, the maximal volume can be written as
	\begin{eqnarray}
		V&=& 2\Omega_{k,d-1}\int^{\infty}_{r_{\rm min}} \frac{\mathrm{d}r}{\dot{r}}\nonumber\\
		&=&2\Omega_{k,d-1} \int^{\infty}_{r_{\rm min}} \mathrm{d}r \, \frac{r^{2(d-1)}}{\sqrt{f(r) r^{2(d-1)} +E^2}} \,.
		\label{volume_int2}
	\end{eqnarray}
	Due to the symmetry of our setting, the point at $r_{\rm min}$ should be a turning point of the surface, and the derivative $\dot{r}$ there should be $0$. Therefore according to Eq.(\ref{rdot}) we have
	\begin{align}
		f(r_{\rm min})\, r_{\rm min}^{2(d-1)} +E^2 = 0 \,.
		\label{eq_rturn}
	\end{align}
	
	Let us consider the value of the coordinate $v$ at the boundary and the turning point, denoted by $v_{\infty}$ and $v_{\rm min}$ respectively. They satisfy the following relation:
	\begin{eqnarray}\label{tau_R}
		v_{\infty}- v_{\rm min} = \tau_{\rm R} + r^{*}(\infty)- r^*(r_{\rm min})\,,
	\end{eqnarray}
	since at the innermost point $t=0$ due to the symmetry.
	Meanwhile, 
	\begin{eqnarray}\label{t_r_E}
		&&v_{\infty}- v_{\rm min}=\int_{v_{\rm min}}^{v_{\infty}} \mathrm{d}v\nonumber\\
		&&=\int^{\infty}_{r_{\rm min}} \mathrm{d}r \left[\frac{E}{f(r)\sqrt{f(r) r^{2(d-1)} +E^2}}+\frac{1}{f(r)}\right]\,.\nonumber\\
	\end{eqnarray}
	
	From Eq.(\ref{tau_R}) and (\ref{t_r_E}) one can derive the time evolution of $\mathrm{d}C_V/\mathrm{d}\tau$.
	Firstly, the time $\tau$ as a function of $r_{\rm min}$ can be written down directly from these two equations and Eq.(\ref{eq_rturn}):
	\begin{eqnarray}\label{tauofrmin}
		\tau &=&2\int^{\infty}_{r_{\rm min}} \mathrm{d}r \frac{E}{f(r)\sqrt{f(r) r^{2(d-1)} +E^2}}\nonumber\\
		&=& -2\int^{\infty}_{r_{\rm min}} \mathrm{d}r \frac{\sqrt{-f(r_{\rm min})}\,  r_{\rm min}^{d-1}}{f(r)\sqrt{f(r) r^{2(d-1)} - f(r_{\rm min})\, r_{\rm min}^{2(d-1)}}}\,.\nonumber\\
	\end{eqnarray}
	The negative sign in the last line is due to the fact that $E$ is negative, which can be see from evaluating Eq.(\ref{tdot}) at $r_{\rm min}$.
	Secondly, one can prove that the following equation holds:
	\begin{eqnarray}
		\frac{V}{2\Omega_{k,d-1}}
		&=&\int^{\infty}_{r_{\rm min}} \mathrm{d}r \,
		\left[\frac{\sqrt{f(r) r^{2(d-1)} +E^2}}{f(r)}+\frac{E}{f(r)}\right] \nonumber\\
		&& - E \left(\tau_{\rm R}+r^*{(\infty)}-r^{*}(r_{\rm min})\right) \,.
	\end{eqnarray}
	Taking the derivative of the above equation with respect to $\tau = \tau_{\rm L}+\tau_{\rm R} = 2\tau_{\rm R}$, one finally arrives at
	\begin{align}
		\frac{\mathrm{d}V}{\mathrm{d}\tau}=\frac{1}{2}\frac{\mathrm{d}V}{\mathrm{d}\tau_{\rm R}} = - \Omega_{k,d-1} E\,,
	\end{align}
	and the growth rate of complexity is
	\begin{equation}
		\frac{d\mathcal{C}_V}{\mathrm{d}\tau} = \frac{1}{G L} \frac{\mathrm{d}V}{\mathrm{d}\tau}
		= \frac{\Omega_{k,d-1}}{G L} \sqrt{-f(r_{\rm min})}\,  r_{\rm min}^{d-1} \, .
		\label{dvdt_general}
	\end{equation}
	
	With Eq.(\ref{tauofrmin}) and (\ref{dvdt_general}) at hand, for any viable value of $r_{\rm min}$ one can find out the boundary time $\tau$ by integration, and get the corresponding value of ${d\mathcal{C}_V}/{\mathrm{d}\tau}$ by algebraic calculation.
	There is one more point.
	To find out the late time limit of $\mathrm{d}C_V/\mathrm{d}\tau$, note that as the total boundary time $\tau$ increases, the maximal surface moves in the future direction, and the turning radius $r_{\rm min}$ decreases. 
	According to Eq.(\ref{rdot}) and (\ref{eq_rturn}), the minimal possible value of $r_{\rm min}$ should be the extreme value point, denoted by $\tilde{r}_{\rm min}$, of the function $\sqrt{-f(r_{\rm min})}r_{\rm min}^{d-1}$. 
	So the late time growth rate is just
	\begin{eqnarray}
		\lim\limits_{\tau\rightarrow \infty}\frac{\mathrm{d} C_V}{\mathrm{d}\tau} = \frac{\Omega_{k,d-1}}{G L} \sqrt{-f(\tilde{r}_{\rm min})}\,  \tilde{r}_{\rm min}^{d-1}\,.
	\end{eqnarray}
	
	Actually the method reviewed here applies to any black hole solution with the form (\ref{staticbh}), in spite of the gravity theory. In the following subsections we first show the results obtained by applying this method for the neutral static AdS black holes with $k=0 ,1$ and $-1$ in EGB gravity, and then we briefly discuss the case of charged black holes with $k=0$.

	\subsection{The Neutral Black Hole with $k=0$}
	In the neutral $k=0$ case, the analytic expression for the late time limit ($\tau\rightarrow\infty$) of $\mathrm{d}C_V/\mathrm{d}\tau$ is
	\begin{eqnarray}\label{latetimek=0}
		&&\lim\limits_{\tau\rightarrow \infty}\frac{\mathrm{d} C_V}{\mathrm{d}\tau}=\frac{2 \sqrt{2} \pi   L M \left(-12 \tilde{\alpha}+L \sqrt{12 \tilde{\alpha}+L^2}+L^2\right) }{(d-1)   \left(L^2-4 \tilde{\alpha}\right)\sqrt{\tilde{\alpha}}}\nonumber\\
		&&\times\sqrt{{\sqrt{\frac{\left(L^2-4 \tilde{\alpha}\right) \left(4 \tilde{\alpha}+L \sqrt{12 \tilde{\alpha}+L^2}+L^2\right)}{L^2 \left(-12 \tilde{\alpha}+L \sqrt{12 \tilde{\alpha}+L^2}+L^2\right)}}-1}}\,,\nonumber\\
	\end{eqnarray}
	and if $\tilde{\alpha}=0$ then we recover the result in Einstein gravity \cite{Stanford:2014jda,Carmi:2017jqz}
	\begin{eqnarray}
		\lim\limits_{\tau\rightarrow \infty}\frac{\mathrm{d} C_V}{\mathrm{d}\tau}=\frac{8\pi M}{d -1 }\,.
	\end{eqnarray}
	The corrections due to the Gauss-Bonnet term are negative, thus suppressing the complexification rate of the black hole. This can be partly seen by looking at the expansion around $\alpha=0$
	\begin{eqnarray}
		\lim\limits_{\tau\rightarrow \infty}\frac{\mathrm{d} C_V}{\mathrm{d}\tau}=\frac{8\pi M}{d -1 }\left(1-\frac{\tilde{\alpha}}{2 L^2} + \frac{11\tilde{\alpha}^2}{8L^4}  + \mathcal{O}((\tilde{\alpha}/L^2)^3)\right)\,.\nonumber\\
	\end{eqnarray}
	For general values of $\tilde{\alpha}$ we find that the late time complexity growth rate is always lower than that in Einstein gravity and decreases as $\tilde{\alpha}$ increases.
	Interestingly, it is always proportional to the black hole mass $M$ as in the case of Einstein gravity.
	Since in the case $k=0$ we have the relation $M\propto S T$, the late time complexity growth rate satisfies the relation
	\begin{eqnarray}\label{ST}
		\lim\limits_{\tau\rightarrow \infty}\frac{\mathrm{d} C_V}{\mathrm{d}\tau}\propto S T\,.
	\end{eqnarray}
	As pointed out in \cite{Stanford:2014jda}, this result is expected based on a quantum circuit model of complexity \cite{Hayden:2007cs,Susskind:2014rva}: the entropy represents the width of the circuit and the temperature is an obvious choice for the local rate at which a particular qubit interacts.

\begin{figure}
		\begin{center}
			\includegraphics[scale=.3]{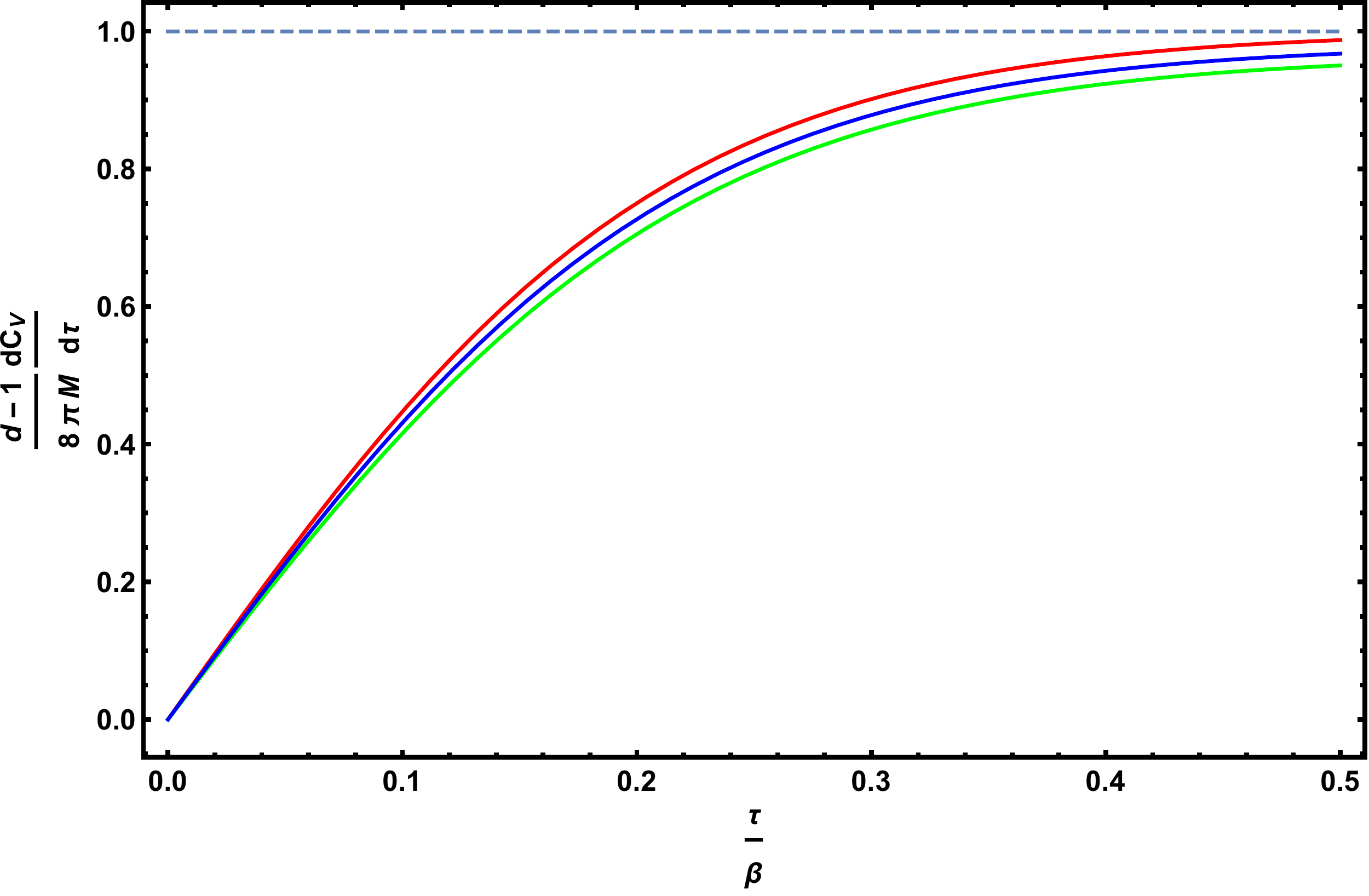}
			\caption{The full time dependence of the complexity growth rate of $5$-dimensional $k=0$ static AdS black holes. 
				The growth rate is converted to a dimensionless quantity by dividing it by $8\pi M/(d-1)$, the late time growth rate in Einstein gravity.
				The horizontal axis is $\tau/\beta$ with $\beta=1/T$, the inverse of the black hole temperature.
				The red curve corresponds to the case in Einstein gravity i.e. $\tilde{\alpha}=0$, and it approaches to $1$ (the dashed line) from below at late times.
				The blue and the green curves correspond to $\tilde{\alpha}/L^2 = 0.04$ and $\tilde{\alpha}/L^2 = 0.08$ in EGB gravity respectively, approaching constant values less than $1$ (corresponding to Eq.(\ref{latetimek=0})) from below at late times.
				In fact the growth rate appears to decrease as the parameter $\tilde{\alpha}/L^2$ increases, at any nonzero time.}\label{fulltime}
		\end{center}
	\end{figure}

	The full time dependence of the complexity growth rate in the case $d=4$ is shown in Figure \ref{fulltime}.
	The growth rate is shown in the dimensionless form $8 \pi (d-1) \mathrm{d}C_V/(M \mathrm{d}\tau)$, proportional to the growth rate per unit energy. In the following while talking about ``complexity growth rate'' we always refer to this dimensionless quantity.
	This dimensionless growth rate in EGB gravity increases monotonically as time goes on, similar to the case of Einstein gravity \cite{Carmi:2017jqz}, while always less than the value in Einstein gravity.
	At $\tau/\beta=0$ the growth rate is $0$, since $r_{\rm min} = r_h$ and the surface does not probe the structure inside the black hole at this time.  
	As $\tau/\beta \rightarrow\infty$ the growth rate approaches a constant which is just the late time value given in Eq.(\ref{latetimek=0}).
	Moreover, the larger the Gauss-Bonnet coupling $\tilde{\alpha}$, the less the growth rate. 
	In fact, if we choose the length scale $\ell$ as
	\begin{eqnarray}
		\ell = \frac{4\pi^2\hbar L}{d-1}
	\end{eqnarray}
	as in the paper \cite{Kim:2017qrq}, the growth rate obeys the Lloyd's bound (\ref{Lloyd}), and saturates it only when the Gauss-Bonnet coupling $\tilde{\alpha}$ vanishes.
	
	\subsection{The Neutral Black Hole with $k=1$}
\begin{figure}
		\begin{center}
			\includegraphics[scale=.3]{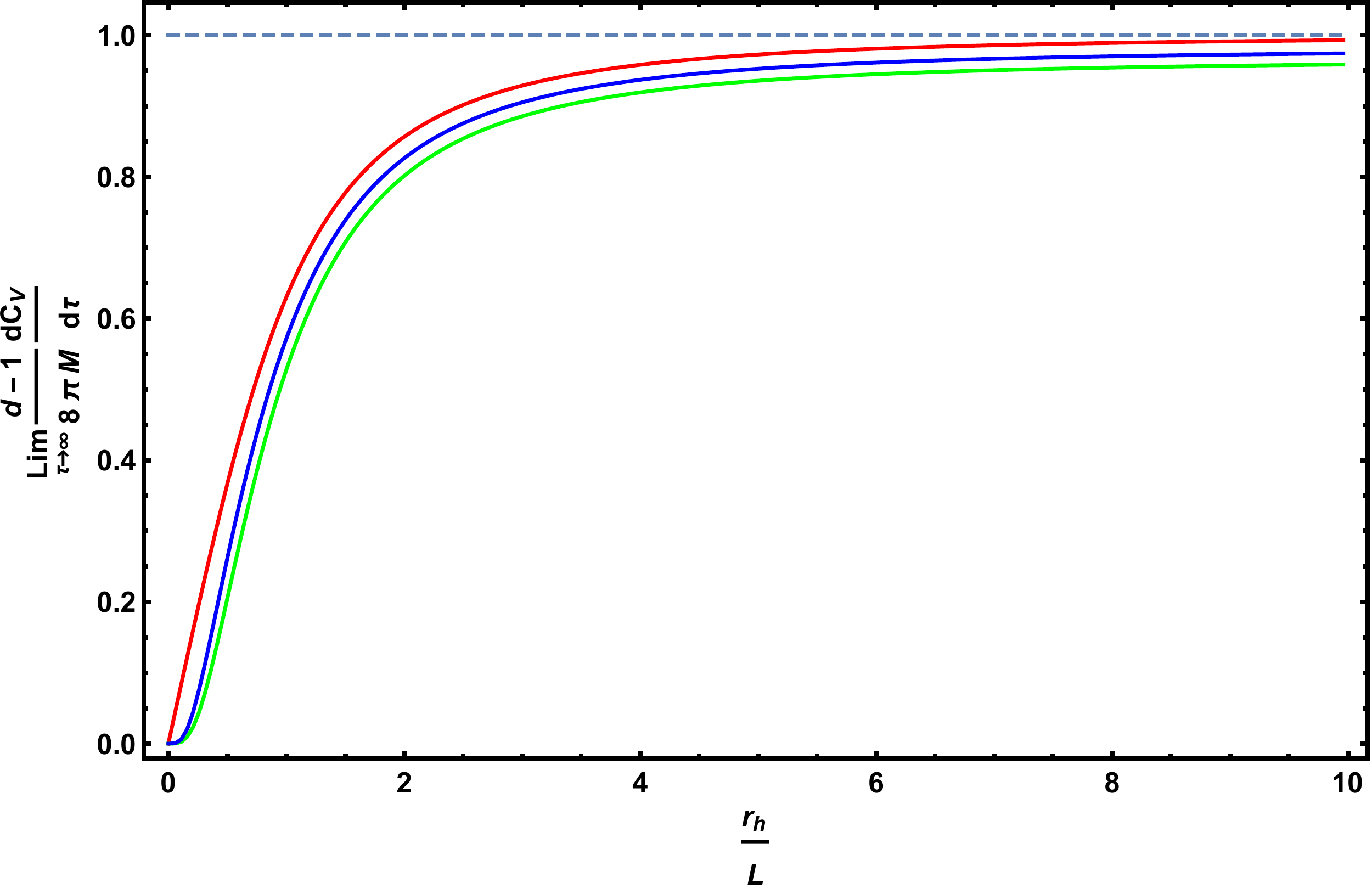}
			\caption{ The late time complexity growth rate as a function of the ratio $r_h/L$ of $5$-dimensional $k=1$ AdS black holes. The red, blue and green curves correspond to the case in Einstein gravity, the cases of $\tilde{\alpha}/L^2 = 0.04$ and $\tilde{\alpha}/L^2 = 0.08$ in EGB gravity respectively. As the size $r_h/L$ of the black hole grows the late time complexity growth rate increases, while approaching to the constant value of the $k=0$ case from below when $r_h/L\rightarrow\infty$. Larger GB couplings correspond to smaller late time growth rates for any nonzero values of $r_h/L$.
			}\label{k=1latetime}
		\end{center}
	\end{figure}

	For the neutral case with $k=  1$, in additional to the Gauss-Bonnet coupling $\tilde{\alpha}$ there is one more parameter describing the black hole --- the radius $r_h$ of the horizon. 
	The analytic expression of the late time complexity growth rate was not found. As an alternative way of presenting the results, its numerical value as a function of the ratio $r_h/L$ for $d=4$ and for different GB couplings are shown in Figure \ref{k=1latetime}. 
	The behavior of the function is similar to that in Einstein gravity.
	The late time limit of the growth rate vanishes when $r_h/L=0$ ( in the absence of black hole horizon), increases as $r_h/L$ increases, and approaches its counterpart in the $k=0$ case in the limit $r_h/L\gg 1$, which can be considered as the high-temperature limit $T L \gg 1$. 
	In this limit, the characteristic thermal wavelength is much shorter than the curvature scale $L$ and the effect of a nonzero $k$ can be ignored. This is also true for the $k=-1$ case in the following subsection.
	Therefore in the high-temperature limit for $k=\pm 1$ black holes the relation (\ref{ST}) is also satisfied. 
	We also find that the larger GB coupling, the smaller late time growth rate, no matter how large $r_h/L$ is. As in the $k=0$ case, the largest late time growth rate is obtained in Einstein gravity.
	
\begin{figure}
		\begin{center}
			\includegraphics[scale=.3]{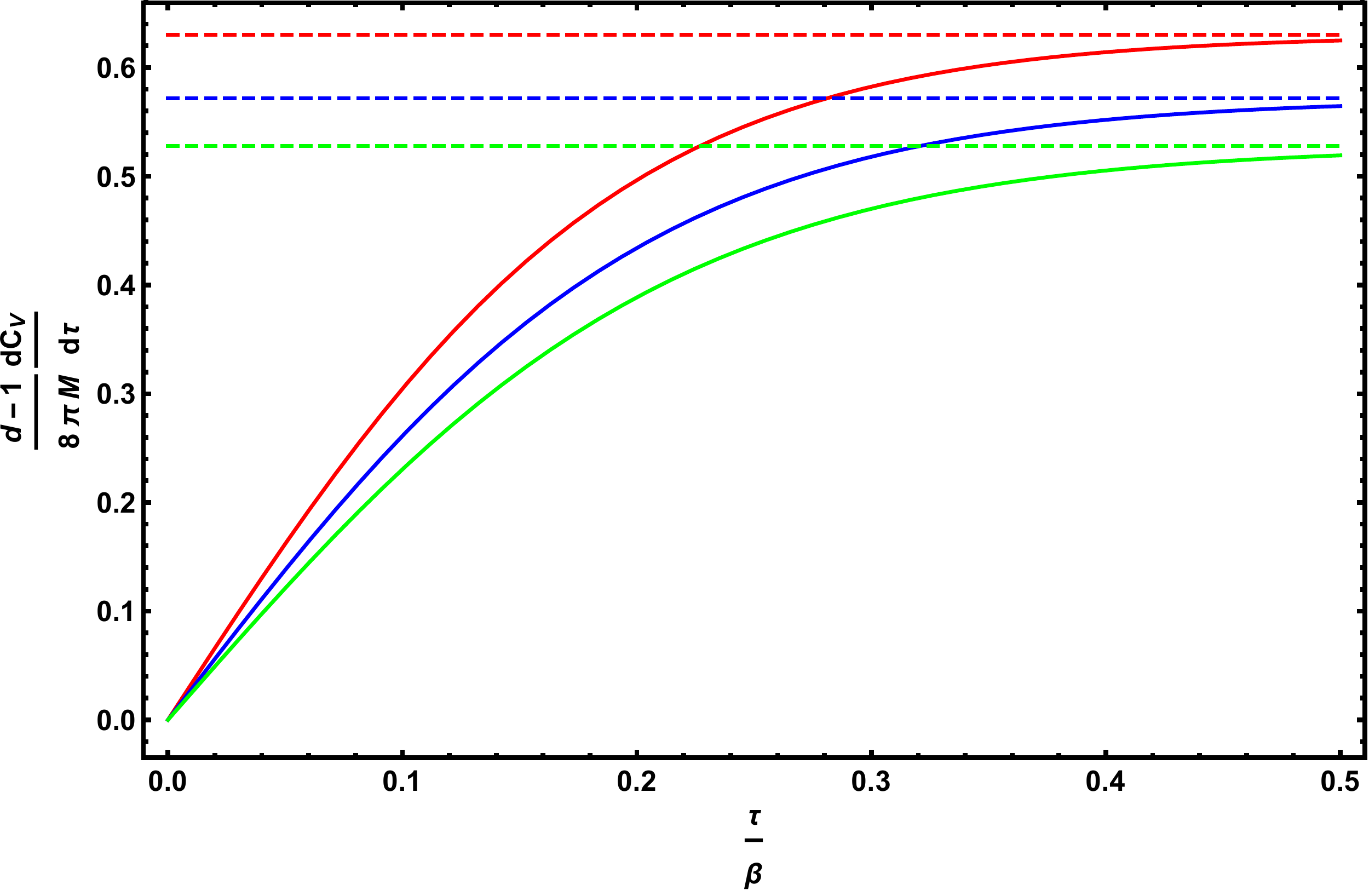}
			\caption{The full time dependence of the complexity growth rate of $5$-dimensional $k=1$ static AdS black holes. The red, blue and green curve correspond to the Einstein gravity case, the cases of $\tilde{\alpha}/L^2 = 0.04$ and $\tilde{\alpha}/L^2 = 0.08$ in EGB gravity respectively. The parameter $r_h/L$ is set to $1$.
				As in the $k=0$ case, each growth rate grows in time and approaches to a constant value from below at late times, while the Gauss-Bonnet coupling suppresses the complexity growth rate.
			}\label{k=1fulltime}
		\end{center}
	\end{figure}

	The full time evolution of the growth rate is similar to that in the planar case. It increases monotonically as time goes by and approaches to the bound from below at late times. Larger Gauss-Bonnet couplings correspond to lower growth rates all the time, as in Einstein gravity.
	This is shown in Figure \ref{k=1fulltime}.
	
	\subsection{The Neutral Black Hole with $k= -1$}
	For the case of a neutral hyperbolic black hole ($k=-1$),
	the mass parameter $M$ can take negative values. In a certain region of the parameter space of $M$, the black hole possesses two horizons, and when the mass takes the minimal allowed value (shown below) which we call $M_{\rm ext}$, the black hole becomes extremal. If $M$ is below this value there is no black hole horizon. 
	The minimal $M$ is
	\begin{eqnarray}
		\label{3eq18}
		M_{\rm ext}&=& -\frac{(d-1)(d-2)\Omega_{k,d-1} L^2 r_{\rm ext}^{d-4}}{16\pi G d^2}\nonumber\\
		&&\times\left ( 1-\frac{d}{d-2} \frac{4\tilde\alpha}{L^2}
		+\sqrt{1-\frac{d(d-4)}{(d-2)^2}\frac{4\tilde\alpha}{L^2}}\right)\,,\nonumber\\
	\end{eqnarray}
	where the minimal value $r_{\rm ext}$ for the parameter $r_h$ satisfies the relation \cite{Cai:2001dz} 
	\begin{equation}
		\label{3eq17}
		r^2_{\rm ext}
		=\frac{(d-2)L^2}{2d}\left( 1+\sqrt{1-\frac{d(d-4)}{(d-2)^2}
			\frac{4\tilde\alpha}{L^2}}\right).
	\end{equation}
	In the case $d=4$ we have
	\begin{eqnarray}
		r_{\rm ext} = \frac{L}{\sqrt{2}}\,,\qquad M_{\rm ext} =\frac{\Omega_{-1,3}  }{64 \pi  G}\left(12 \tilde{\alpha}-3 L^2\right)\,.
	\end{eqnarray}
	As has been done in \cite{Carmi:2017jqz}, to avoid negative energy, in the $k=-1$ case we calculate the following quantity as the dimensionless version of the complexity growth rate:
	\begin{eqnarray}
		\frac{d-1}{8\pi (M- M_{\rm ext})} \frac{\mathrm{d} C_V}{\mathrm{d}\tau}\,.
	\end{eqnarray}
	For different values of $M$, there can be an inner horizon as well as a singularity between the outer horizon $r=r_h$ and $r=0$.
	However, the maximal surface never reaches the inner horizon or the singularity\footnote{This is because that once the turning point $r_{\rm min}$ exists, it is definitely outside any singularity, and it should also be in the region where $f(r)<0$ (outside the inner horizon) due to the embedding equations describing the maximal surface.}. 
	
\begin{figure}
		\begin{center}
			\includegraphics[scale=.3]{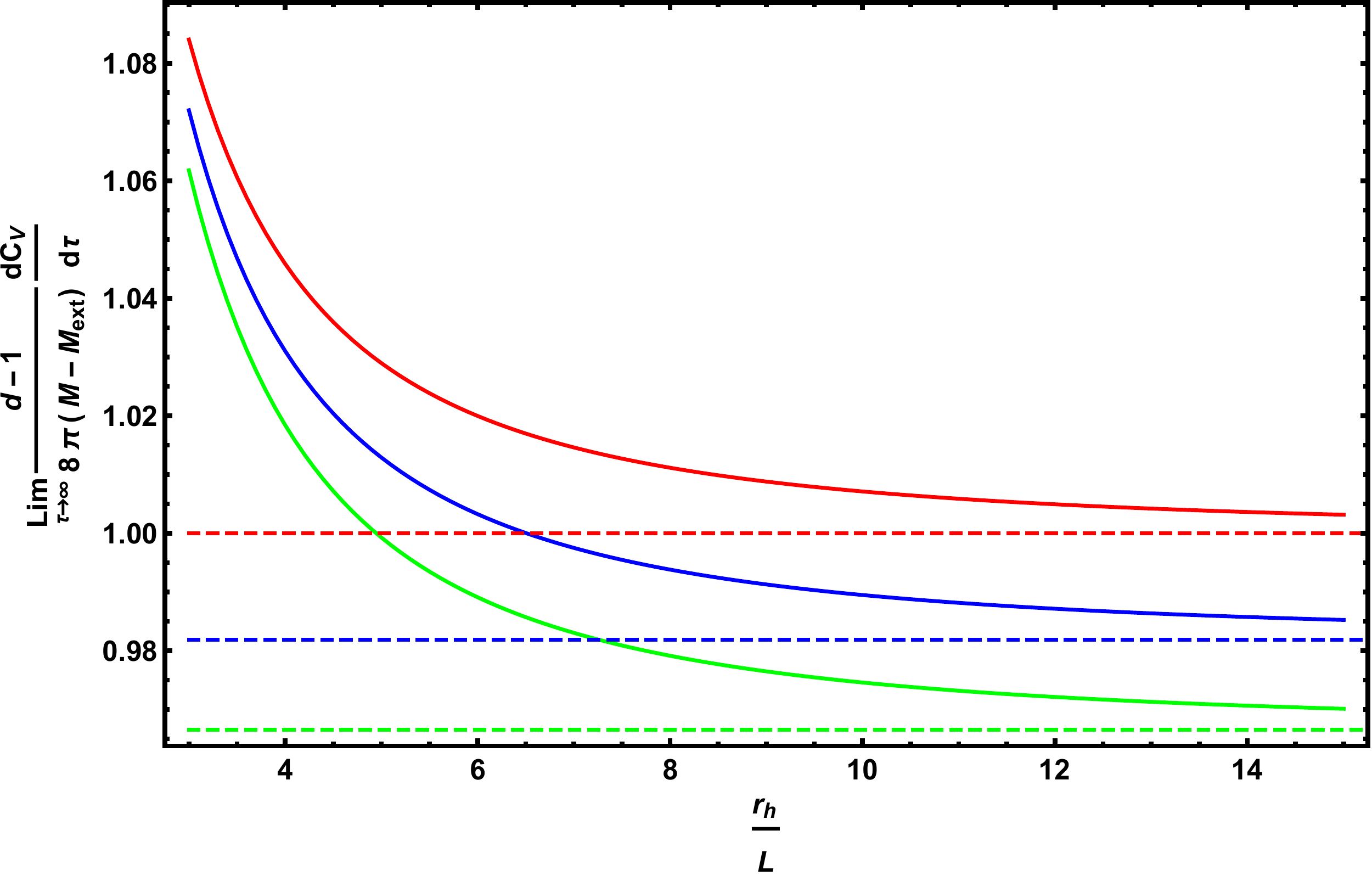}
			\caption{ The late time complexity growth rate as a function of the ratio $r_h/L$ ($r_h/L \geq 3$) of $5$-dimensional $k=-1$ AdS black holes.
				The red, blue and green curves correspond to the case in Einstein gravity, the cases of $\tilde{\alpha}/L^2 = 0.04$ and $\tilde{\alpha}/L^2 = 0.08$ in EGB gravity respectively.
				As the size $r_h/L$ of the black hole grows the late time complexity growth rate decreases, while approaching to the constant value of the $k=0$ case from above when $r_h/L\rightarrow\infty$. In the range in this figure larger GB couplings correspond to smaller late time growth rates.
			}\label{k=-1latetime}
		\end{center}
	\end{figure}

\begin{figure}
		\begin{center}
			\includegraphics[scale=.3]{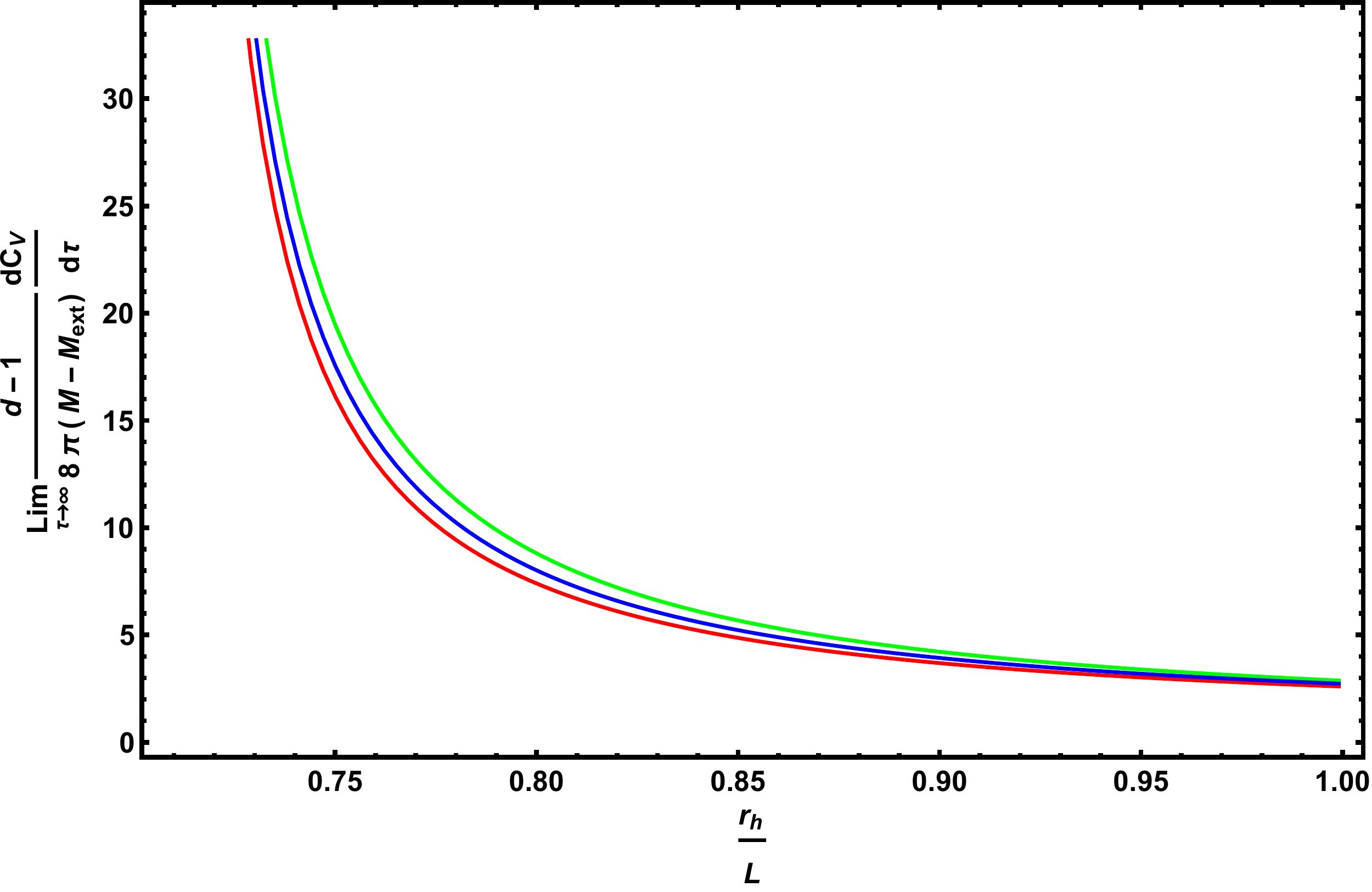}
			\caption{ The late time complexity growth rate as a function of the ratio $r_h/L$ ($r_{\rm ext}\leq r_h\leq L$) of $5$-dimensional $k=-1$ AdS black holes.
				The red, blue and green curves correspond to the case in Einstein gravity, the cases of $\tilde{\alpha}/L^2 = 0.04$ and $\tilde{\alpha}/L^2 = 0.08$ in EGB gravity respectively. 
				In the parameter range in this figure larger GB couplings correspond to larger late time growth rates.}\label{k=-1latetime2}
		\end{center}
	\end{figure}

	The late time limit of $(d-1)/(8\pi (M- M_{\rm ext})) {\mathrm{d} C_V}/{\mathrm{d}\tau}$ as a function of the ratio $r_h/L$ is plotted in Figure \ref{k=-1latetime} and Figure \ref{k=-1latetime2}.
	The function goes to infinity as $r_h/L$ approaches $r_{\rm ext}/L$, and decreases as $r_h/L$ increases. Similar to the $k=1$ case, it also approaches the late time growth rate in the $k=0$ case, however from above.
	So in this case the Lloyd's bound is not obeyed. However for certain values of $\alpha/L^2$ and $r_h/L$ there is always an upper bound in the time evolution, as will be shown later.
	If $r_h/L$ is not too small, a larger Gauss-Bonnet coupling corresponds to a smaller late time complexity growth rate. This fact is shown in Figure \ref{k=-1latetime}.
	However, for very small values of $r_h/L$, a larger Gauss-Bonnet coupling corresponds to a larger late time growth rate, as shown in Figure \ref{k=-1latetime2}.
	In this case the complexity growth rate in Einstein gravity is the smallest.
	This fact is in contrast to the flat and spherical cases, where the growth rate in Einstein gravity is always the largest.
	
\begin{figure}
		\begin{center}
			\includegraphics[scale=.3]{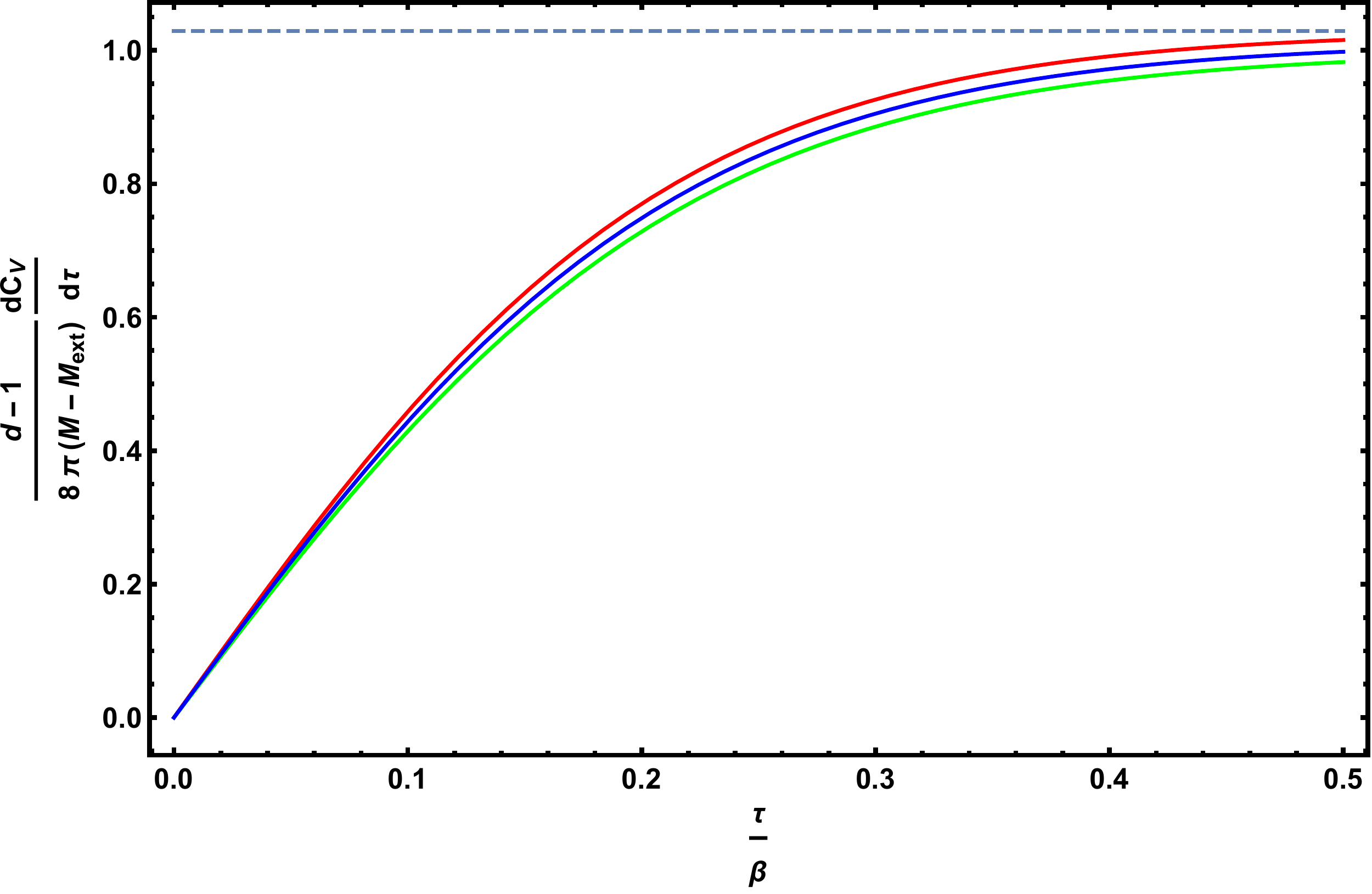}
			\caption{The full time dependence of the complexity growth rate of $5$-dimensional $k=-1$ static AdS black holes.
				The red, blue and green curves correspond to the case in Einstein gravity, the cases of $\tilde{\alpha}/L^2 = 0.04$ and $\tilde{\alpha}/L^2 = 0.08$ in EGB gravity respectively.	
				The parameter $r_h/L$ is set to $5$, and the Gauss-Bonnet coupling suppresses the complexity growth rate.
			}\label{k=-1fulltime}
		\end{center}
	\end{figure}

\begin{figure}
		\begin{center}
			\includegraphics[scale=.3]{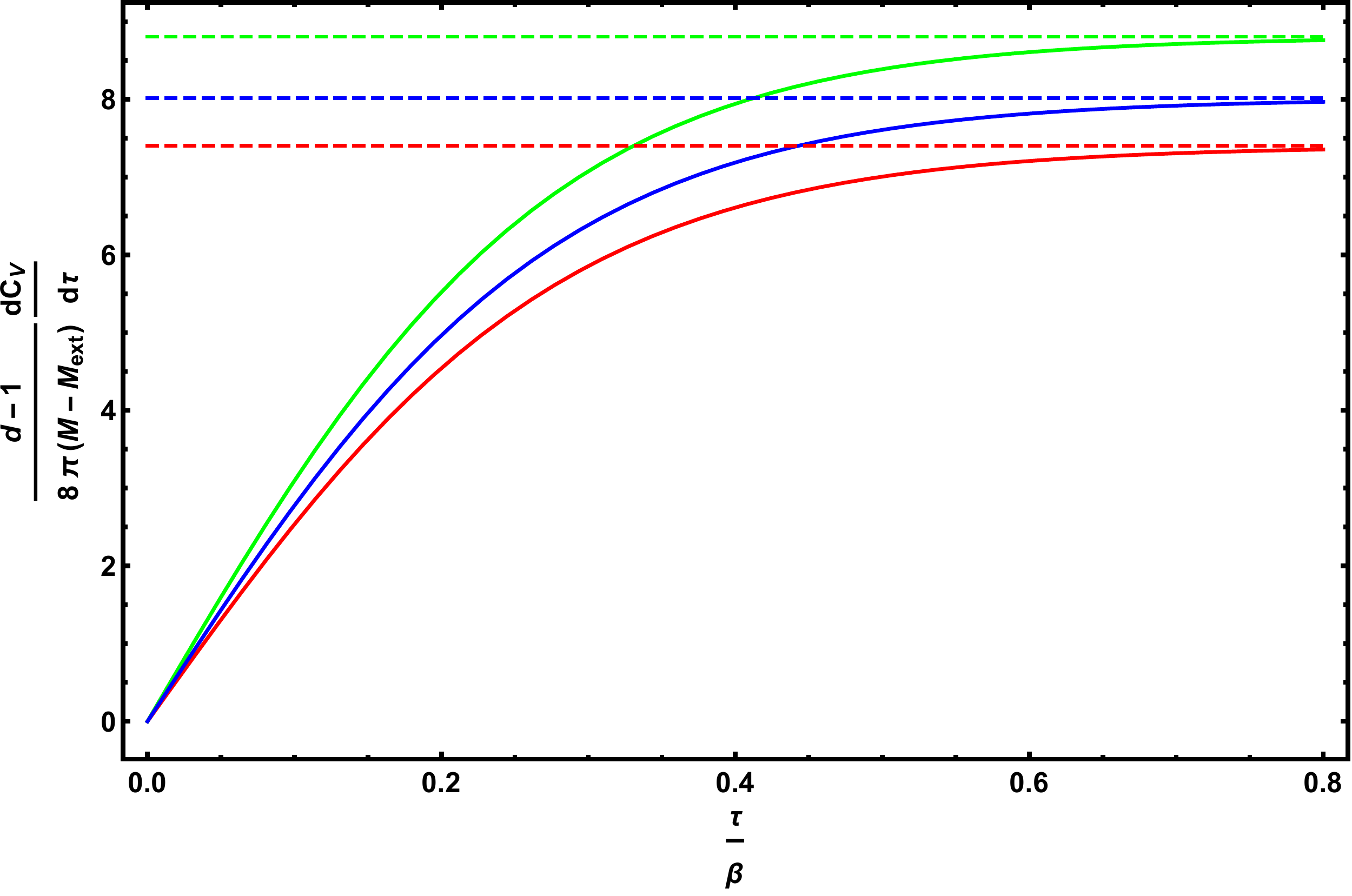}
			\caption{The full time dependence of the complexity growth rate of $5$-dimensional $k=-1$ static AdS black holes.
				The red, blue and green curves correspond to the case in Einstein gravity, the cases of $\tilde{\alpha}/L^2 = 0.04$ and $\tilde{\alpha}/L^2 = 0.08$ in EGB gravity respectively.
				The parameter $r_h/L$ is set to $0.8$, and the Gauss-Bonnet coupling enhances the complexity growth rate.
			}\label{k=-1fulltime2}
		\end{center}
	\end{figure}

	Figure \ref{k=-1fulltime} shows the full time dependence of the complexity growth rate of larger black holes ($r_h/L=5$), while Figure \ref{k=-1fulltime2} shows the results of relatively small black holes ($r_h/L=0.8$).
	In the former case, a larger Gauss-Bonnet coupling corresponds to a smaller complexity growth rate in full time. The latter case is the opposite --- a larger Gauss-Bonnet coupling corresponds to a larger growth rate in full time.
	In both cases the growth rate increases monotonically in time and approaches the late time limit from below.
	
	\subsection{The Charged Black Hole with $k=0$}

\begin{figure}
		\begin{center}
			\includegraphics[scale=.3]{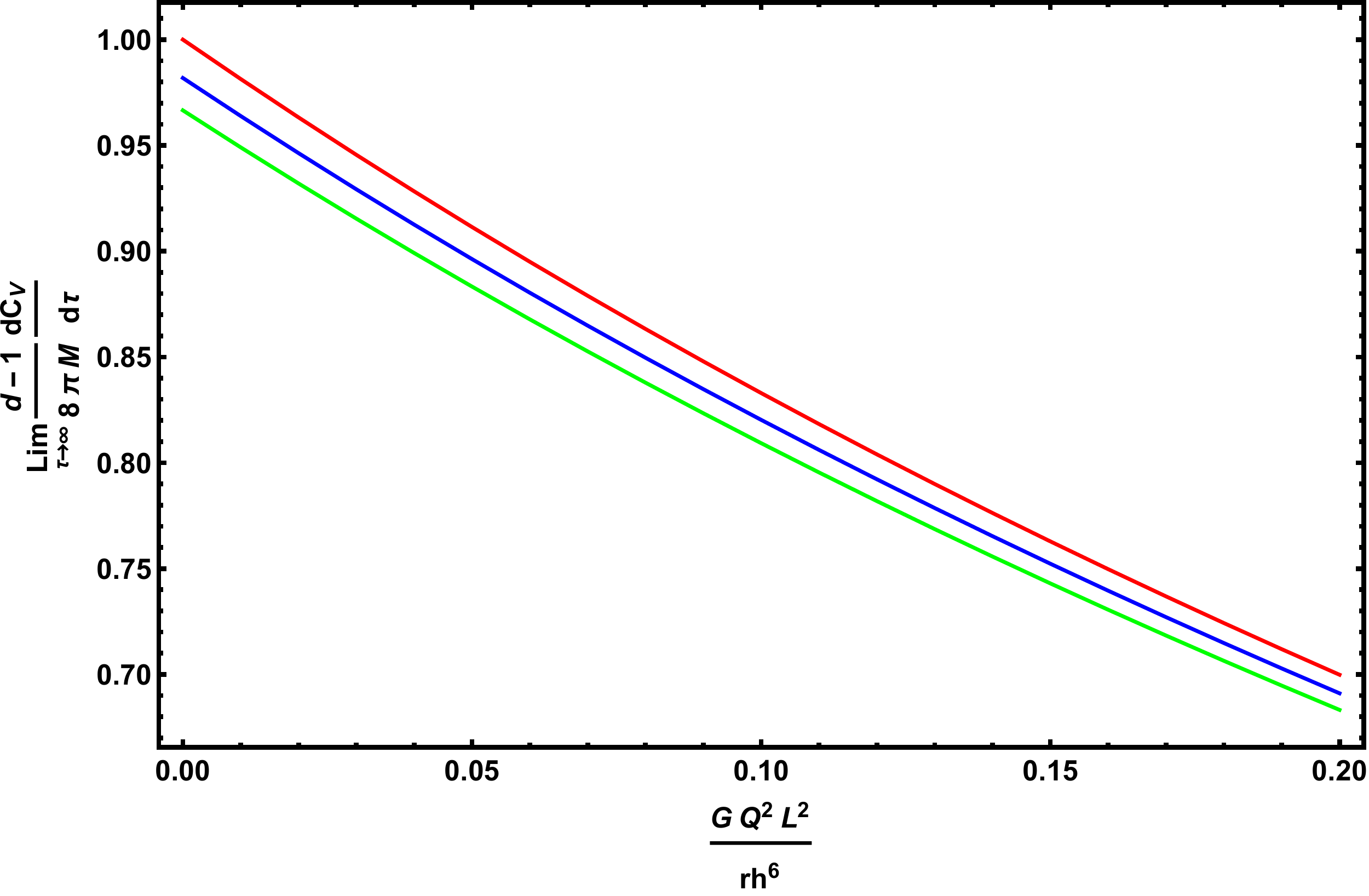}
			\caption{ The late time complexity growth rate as a function of the ratio $G Q^2 L^2/r_h^6$, representing the magnitude of the electric charge, of $5$-dimensional charged planar AdS black holes.
				The red, blue and green curves correspond to the case in Einstein gravity, the cases of $\tilde{\alpha}/L^2 = 0.04$ and $\tilde{\alpha}/L^2 = 0.08$ in EGB gravity respectively. As the charge of the black hole grows, the late time complexity growth rate decreases. The growth rate continuously approaches the value of the neutral black hole when the charge goes to zero. Though not shown in this figure, it reaches zero when the charge reaches the maximum value corresponding to the extremal black hole. The Gauss-Bonnet parameter suppresses the complexity growth rate.
			}\label{chargedlatetime}
		\end{center}
	\end{figure}

	This subsection will briefly  discuss the effect of electric charge of the black hole to the complexity growth rate. For simplicity we consider the $k=0$ case. Unlike in the CA conjecture \cite{Brown:2015lvg,Cai:2016xho} where the analytic expression is present, we only obtained the numerical value of the growth rate as a function of the charge or time.
	Figure \ref{chargedlatetime} shows the late time complexity growth rate as a function of the charge of the black hole.
	The growth rate decreases as the charge increases, reaching zero for the extremal black hole, while the limit of zero charge reproduces the result in the neutral case. This is similar to the results in Einstein gravity found in \cite{Carmi:2017jqz}.
	Moreover, we find that as in the neutral case, the larger the Gauss-Bonnet parameter, the lower the growth rate. In a word, both the conserved charge and the Gauss-Bonnet coupling suppress the rate of complexification.
	
\begin{figure}
		\begin{center}
			\includegraphics[scale=.3]{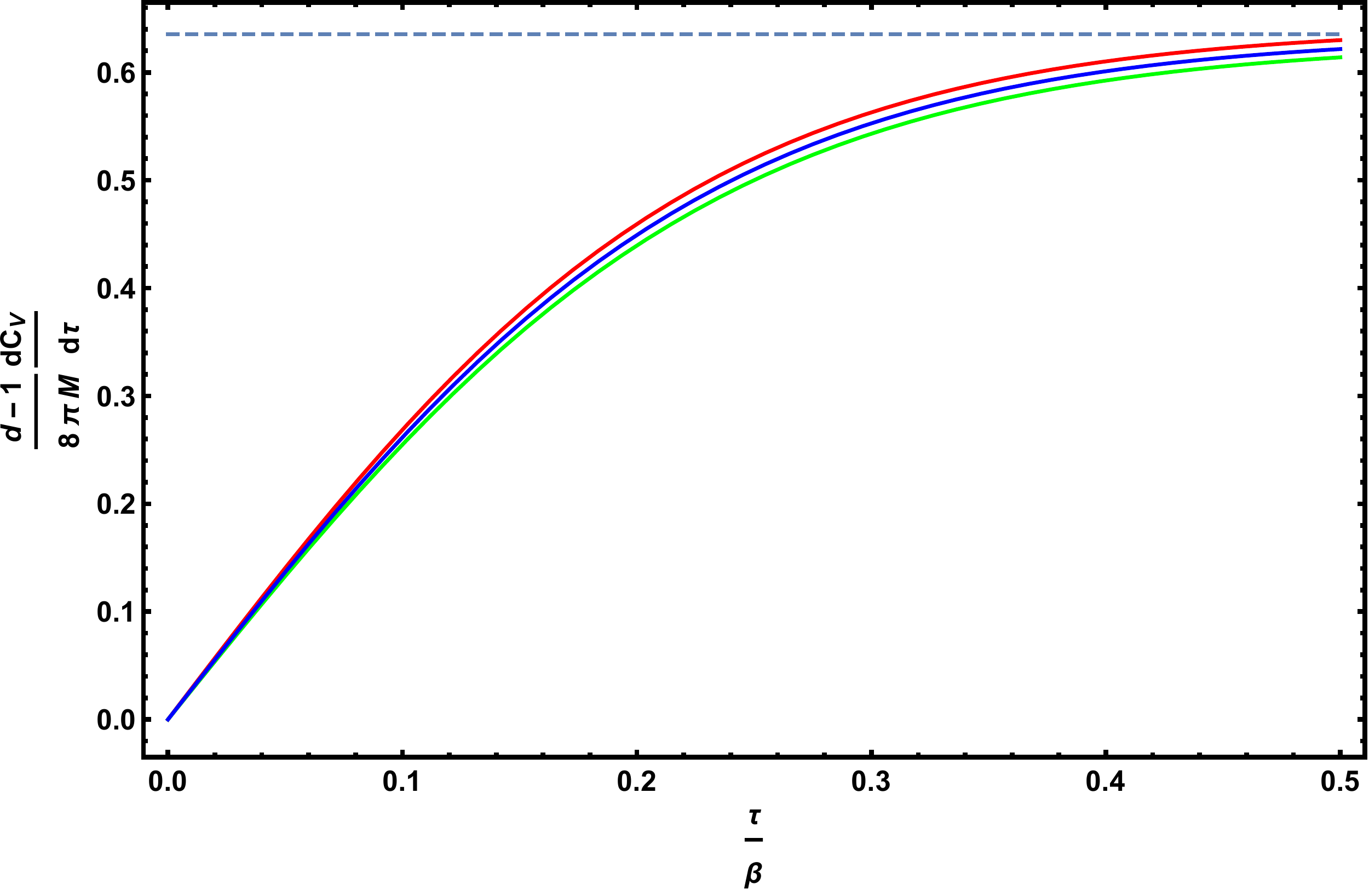}
			\caption{The full time dependence of the complexity growth rate of $5$-dimensional charged planar AdS black holes.
				The red, blue and green curves correspond to the case in Einstein gravity, the cases of $\tilde{\alpha}/L^2 = 0.04$ and $\tilde{\alpha}/L^2 = 0.08$ in EGB gravity respectively.  The parameter $Q$ is set to satisfy
				$G Q^2 L^2/r_h^6 = 0.25$. As in the neutral case, each growth rate grows monotonically in time and approaches to a constant value from below at late times.
			}\label{chargedfulltime}
		\end{center}
	\end{figure}

	According to the numerical results for $d=4$ shown in Figure \ref{chargedfulltime}, the full time dependence of the complexity growth rate behaves in a similar way as the neutral black hole. The upper bound is just the late time value of the monotonically increasing growth rate.
	In a word, the effect of the Gauss-Bonnet coupling is not very much different from that of the neutral case.

  	\section{The Growth Rate of Complexity in the CV2.0 Conjecture}\label{CV2}
	\subsection{Late Time Results}

	The CV2.0 conjecture was first proposed in \cite{Couch:2016exn}  by analyzing the relation between complexity and black hole chemistry, and it says that the holographic complexity is proportional to the spacetime volume of the WdW patch. 
	The conjecture is formulated as below
	\begin{eqnarray}
		C_V=\frac{1}{\hbar} P\times V_{\rm WdW}\,,
	\end{eqnarray}
	where $P$ is the pressure, and $V_{\rm WdW}$ is the spacetime volume of the WdW patch.
	It is interesting to see if we can obtain similar results as in the previous section in this conjecture.
	In this section we only study the neutral black holes.
	In various examples of  \cite{Couch:2016exn}, it was shown that the late time complexity growth rate is expressed as
	\begin{eqnarray}
		\lim\limits_{\tau\rightarrow \infty}\frac{\mathrm{d} C_V}{\mathrm{d}\tau}=\frac{PV_{\rm th}}{\hbar}\,,
	\end{eqnarray}
	where $V_{\rm th}$ is the so-called thermodynamic volume. 
	For the Schwarzschild-AdS black hole in Einstein gravity, the result is 
	\begin{equation}
		\lim\limits_{\tau\rightarrow \infty}\frac{\mathrm{d} C_V}{\mathrm{d}\tau}=\frac{M}{\hbar}\,.
	\end{equation}
	
	As discussed in Ref.\cite{Carmi:2017jqz}, the complexity does not increase until the critical time, where the critical time is denoted by
	\begin{equation} \label{fdfsfsf}
		\tau_{c}=2(r^{*}(\infty)-r^{*}(r_{s}))\,.
	\end{equation}
	We denote $r_{s}$ the position of singularity, and in the $k=0$ and $k=1$ cases, the singularity is located at $r=0$, while for $k=-1$ there is an additional singularity at $r_{s}>0$ apart from $r=0$ when $M_{ext}<M<0$. The location of singularity is \cite{Cai:2001dz}
	\begin{equation}
		r_{s}^{d}=\frac{4\tilde{\alpha} r_{h}^{d-2}}{1-4\tilde{\alpha}/L^{2}}(1-\frac{\tilde{\alpha}}{r_{h}^{2}}-\frac{r_{h}^{2}}{L^{2}})\,.
	\end{equation}
	
	When the time is above the critical time, to concretely calculate the spacetime volume growth of the Gauss-Bonnet black hole, we partition the WdW patch into several parts.
	We just need to calculate the volumes of the regions $1, 2$ and $3$ ($V_1, V_2$ and $V_3$) in Figure \ref{CAPenrose}, and multiply the sum of these volumes by $2$.
	The integration formula of the volume is
	\begin{eqnarray}
		V_{\rm WdW}=\int_{\rm WdW} d^{d+1} x \sqrt{-g}=\Omega_{k,d-1} \int \mathrm{d}t \mathrm{d}r r^{d-1}\,.
	\end{eqnarray}
	Consider the case in which $r=r_{s}$ is the singularity, and
	\begin{eqnarray}
		V_{1}= \Omega_{k,d-1} \int_{r_{s}} ^{r_h} \mathrm{d}r r^{d-1}( \tau_{\rm R}+r^{*}(\infty)-r^{*}(r))
	\end{eqnarray}
	
	\begin{eqnarray}
		V_{2}=2 \Omega_{k,d-1} \int_{r_h} ^{\infty} \mathrm{d}r  r^{d-1}( r^{*}(\infty)-r^{*}(r))
	\end{eqnarray}
	
	\begin{eqnarray}
		V_{3}= \Omega_{k,d-1} \int_{r_{\rm m}} ^{r_{h}} \mathrm{d}r r^{d-1}( -\tau_{\rm R}+r^{*}(\infty)-r^{*}(r))
	\end{eqnarray}
	
	So 
	\begin{eqnarray}
		\frac{\mathrm{d} V_{\rm WdW}}{\mathrm{d}\tau}=\frac{ \Omega_{k,d-1}}{d}( r_{\rm m}^{d}-r_{s}^{d})\,,
	\end{eqnarray}
	where $r_{\rm m}$ is calculated by the equation 
	\begin{eqnarray}
		\frac{\tau-\tau_{c}}{2}=r^{*}(r_{s})-r^{*}(r_{\rm m})\,.
		\label{solvrm}
	\end{eqnarray}
	Except for case $k=-1$ and $M_{ext}<M<0$ , there is only one spacetime singularity at $r=0$. 
	We consider the case $r_{s}=0$ in the following and will discuss the case $ r_{s} \neq 0 $ separately.
	In the late time limit $r_{\rm m} \to r_{h}$, recall that in extended phase space pressure is identified as the cosmological constant $P=-{\Lambda}/{8\pi}$, we find that the complexity growth rate is 
	\begin{eqnarray}
		\frac{\mathrm{d} C_V}{\mathrm{d}\tau}=\frac{(d-1)\Omega_{k,d-1} r_{h}^d }{16 \pi L^{2}}\,.
	\end{eqnarray}
	As is calculated in Ref. \cite{Cai:2013qga}, what appears on the right hand side is precisely the thermodynamic volume 
	\begin{eqnarray}
		V_{\rm th}=(\frac{\partial H}{\partial P})_{S}=\frac{\Omega_{k,d-1} r_{h}^{d}}{d}\,.
	\end{eqnarray}
	
	As to the effect of the higher curvature terms on the holographic complexity growth, invoking the mass expression of the Gauss-Bonnet black hole is a direct way to find it out. 
	For $k=0$ case, we find that the complexity growth rate is independent of higher curvature corrections, which is different from the CV conjecture.  It follows from that the thermodynamic relations of $k=0$ GB black hole are the same as those of the Einstein black hole despite very different geometries.  
	
\begin{figure}
		\includegraphics[scale=.375]{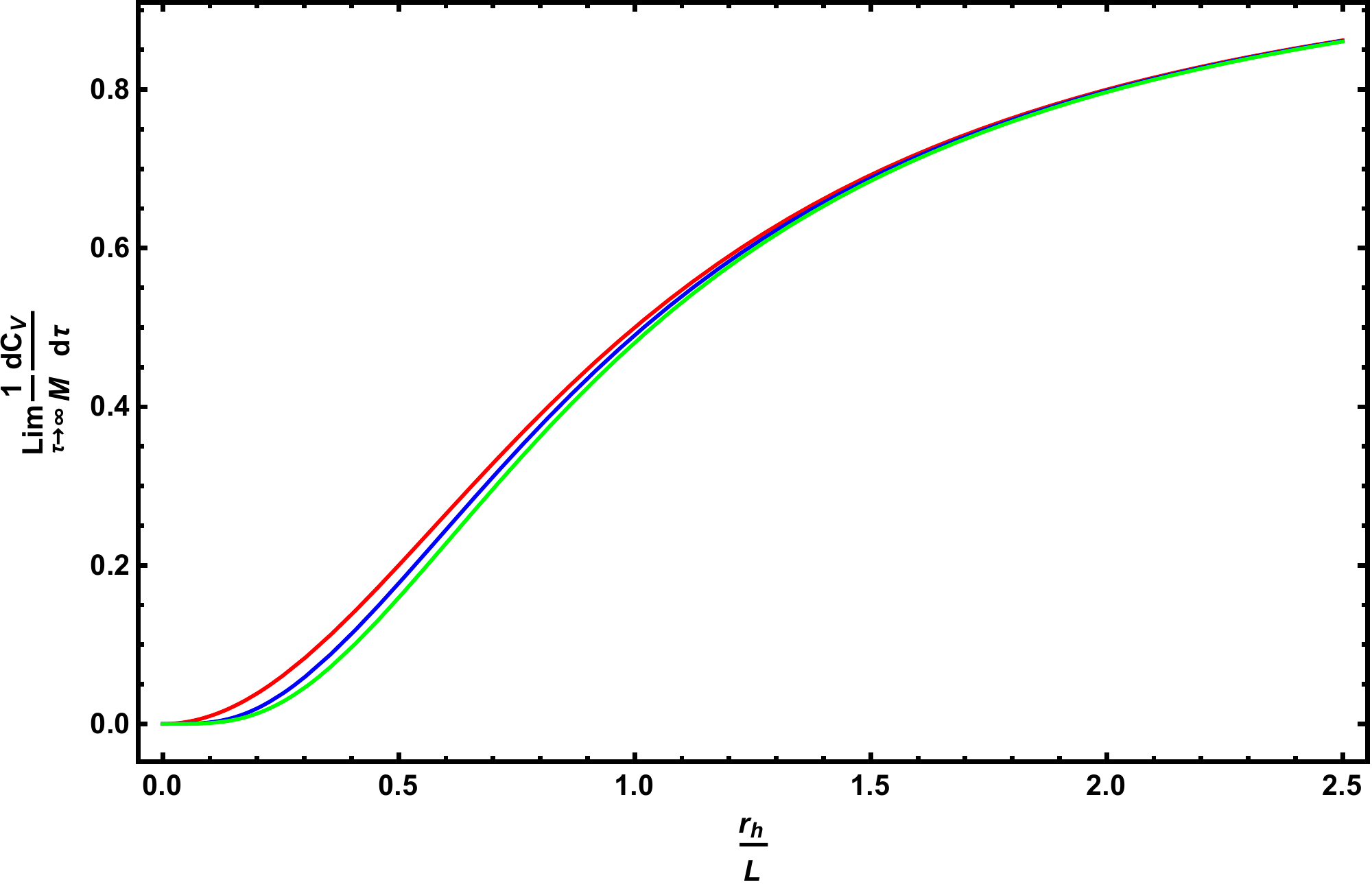}\\
		\caption{Relation between the horizon radius and complexity growth rate in late time limit for the $k=1, d=4$ case. The red curve is the Einstein gravity, the blue curve is $\tilde{\alpha}=0.04 L^{2}$, and the green curve is $\tilde{\alpha}=0.08 L^{2}$. The Gauss-Bonnet coupling suppresses the complexity growth.}	\label{gggg}
	\end{figure}

\begin{figure}
		\includegraphics[scale=.375]{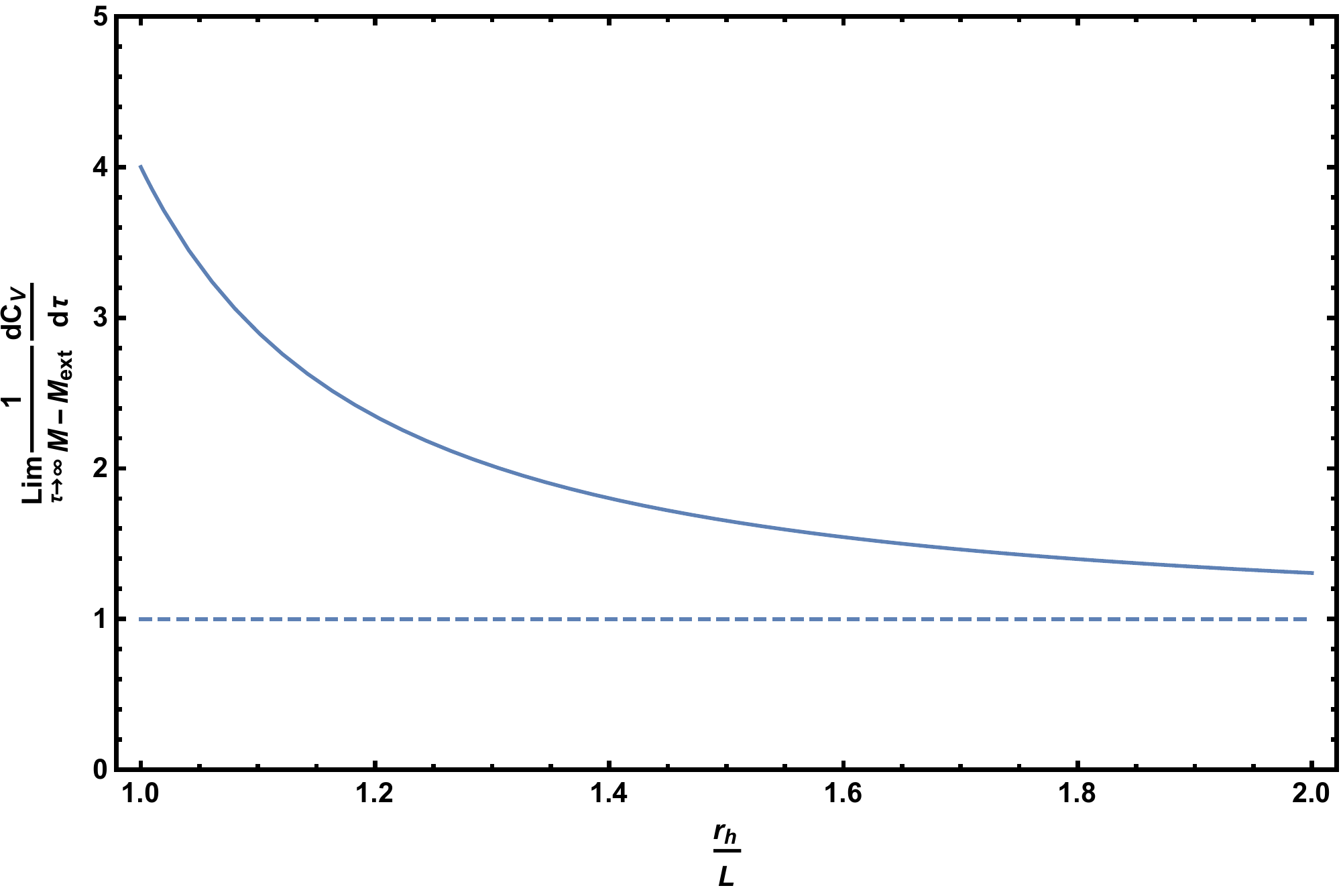}\\
		\caption{Relation between the horizon radius and complexity growth rate in late time limit for the $k=-1, d=4$ case. We consider the region of the radius corresponding to $M>0$, and note that we find exact cancellation of the effects of $\tilde{\alpha}$, so the Gauss-Bonnet coupling does not affect the late time results.}	\label{isfjsf}
	\end{figure}

	For $k \neq 0$ cases, the result is 
	\begin{eqnarray}
		\frac{\mathrm{d} C_V}{\mathrm{d}\tau}=M-\frac{(d-1)\Omega_{k,d-1} r_h^{d-2}} {16 \pi G} (k+\frac{k^2\tilde{\alpha}}{r_h^{2}})\,.
	\end{eqnarray}
	The late time results of the complexity growth rate as functions of the black hole size $r_h/L$ are shown in Figure \ref{gggg} for $k=1$ and Figure \ref{isfjsf} for $k=-1$.

	\subsection{Full Time Behavior}
\begin{figure}
		\begin{center}
			\includegraphics[scale=.375]{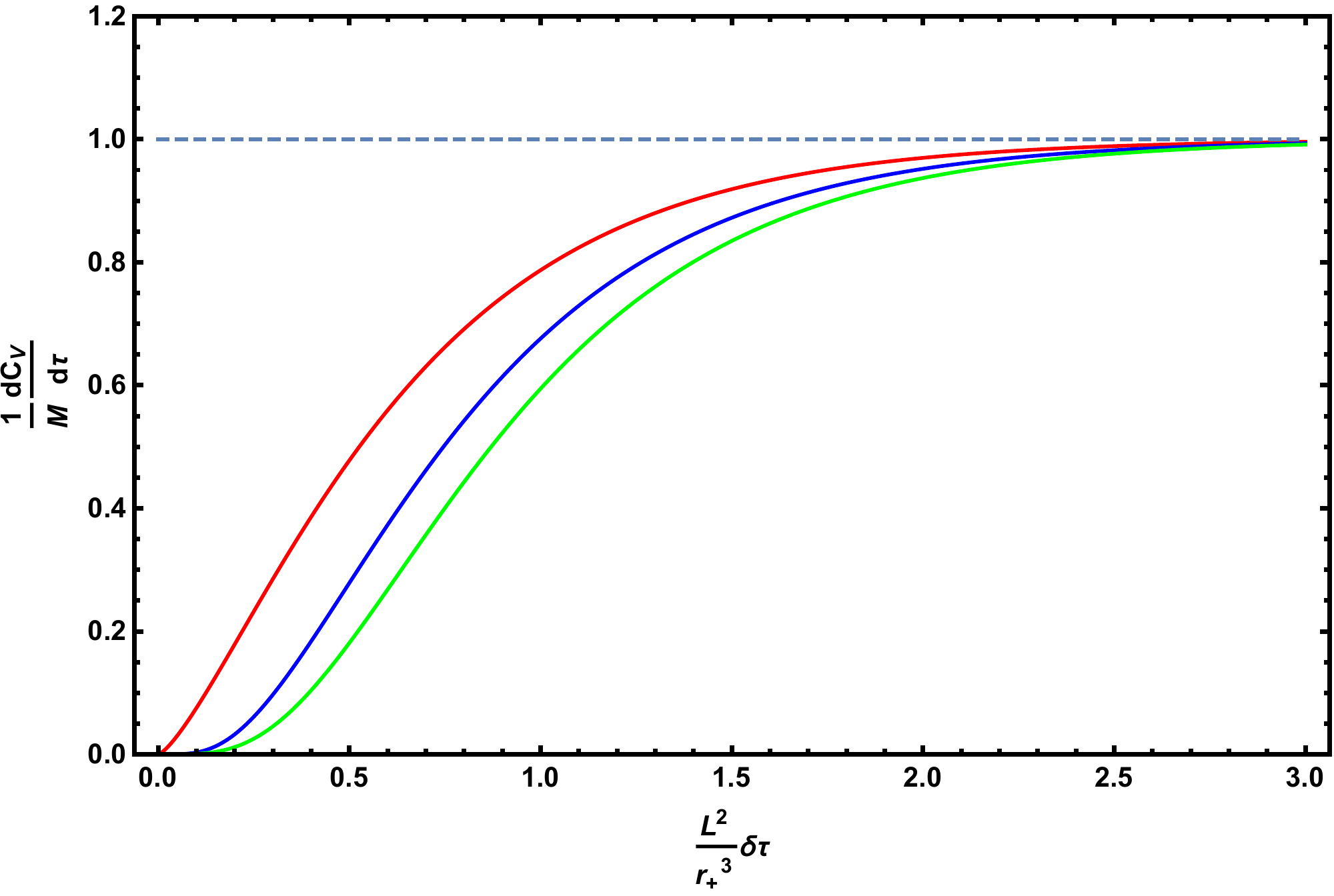}\\
			\caption{Full time dependence of complexity growth rate when $k=0$. The red curve is Einstein case, and the blue curve the $\tilde{\alpha}=0.04 L^{2}$ case, the green curve the $\tilde{\alpha}=0.08 L^{2}$ case, respectively.}\label{dddd}
		\end{center}	
	\end{figure}

	The full time evolution of complexity is easily performed by first solving the equation ($\ref{solvrm}$) to get $r_{\rm m}$, and then plugging it in the full time result of the complexity growth rate
	\begin{eqnarray}
		\frac{\mathrm{d} C_V}{\mathrm{d}\tau}=\frac{(d-1)\Omega_{d-1} r_{\rm m}^d }{16 \pi L^{2}}\,.
	\end{eqnarray}
	Using the expression of $M$ (\ref{M}), for the simplest case $k=0$ we get the full time dependence of complexity growth rate in Figure \ref{dddd}.
	We see that while the Gauss-Bonnet coupling does not change the late time result, it reduces the full time value of the complexity growth rate. We also find that the results in the CV2.0 conjecture is similar to the CV conjecture, where they both grow monotonically. 
	
\begin{figure}
		\includegraphics[scale=.375]{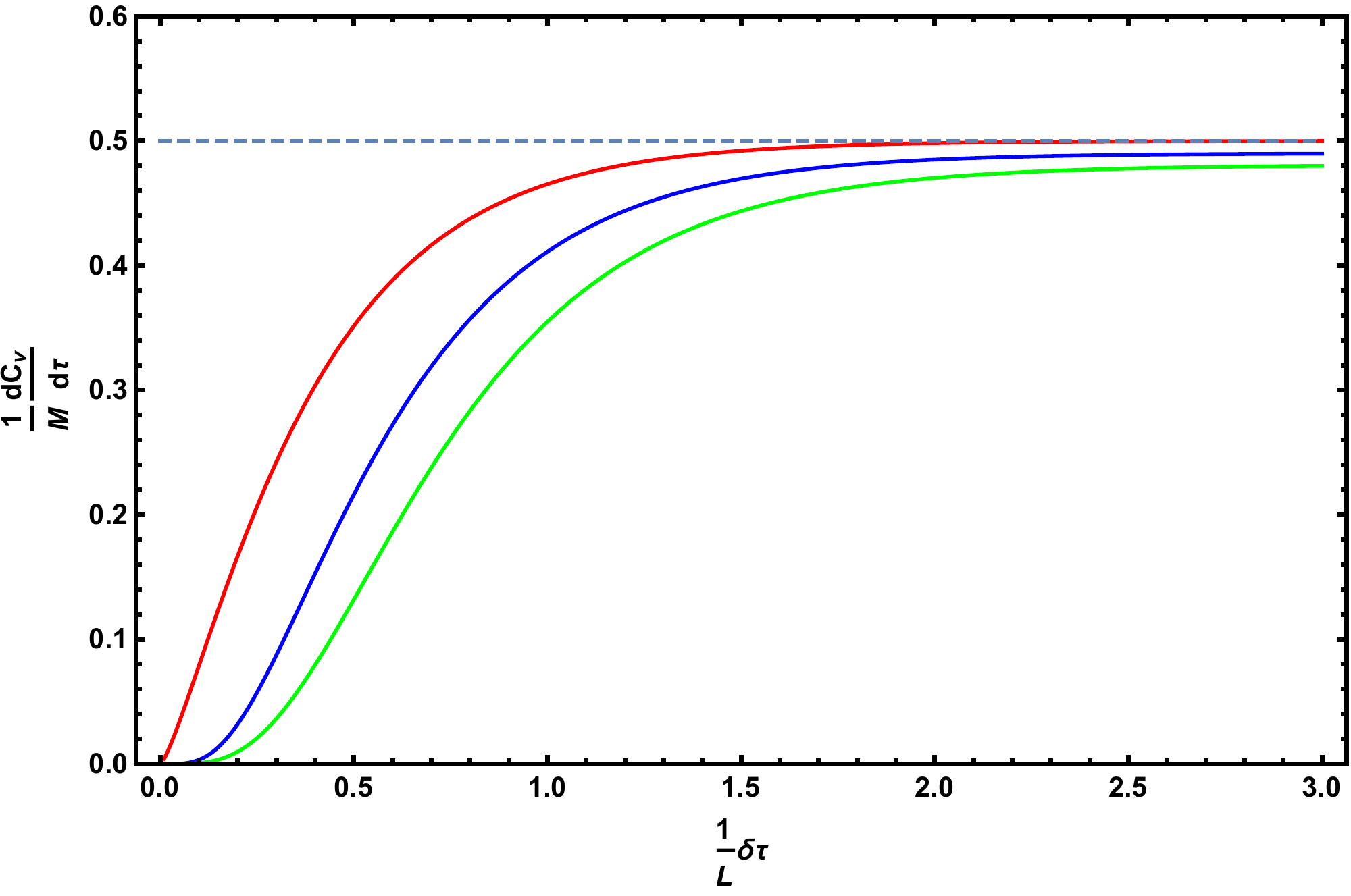}\\
		\caption{Full time dependence of complexity growth rate when k=1. The red curve is Einstein gravity, and blue curve $\tilde{\alpha}=0.04 L^{2}$, green curve $\tilde{\alpha}=0.08 L^{2}$, respectively, and we fix $b=1$ here. }	\label{ffff}
	\end{figure}

	For the $k=1$ case, the expression of the complexity growth rate at a general time is as follows,
	\begin{eqnarray}
		\frac{\mathrm{d} C_V}{\mathrm{d}\tau}=M (\frac{r_{\rm m}}{r_h})^{d} \frac{b^{2}}{1+\frac{a}{b^{2}}+b^{2}}\,,
	\end{eqnarray}
	where $b\equiv r_h/L$ and $a\equiv{\tilde{\alpha}}/{L^{2}}$.
	We plot the full time evolution of complexity for fixed $b$ in Figure \ref{ffff} and find that it behaves qualitatively the same as the $k=0$ case, except that the late time limit depends on the value of $\tilde{\alpha}$. In this case, the presence of the nonzero $\tilde{\alpha}$ always decreases the complexity growth rate. 

\begin{figure}
		\includegraphics[scale=.375]{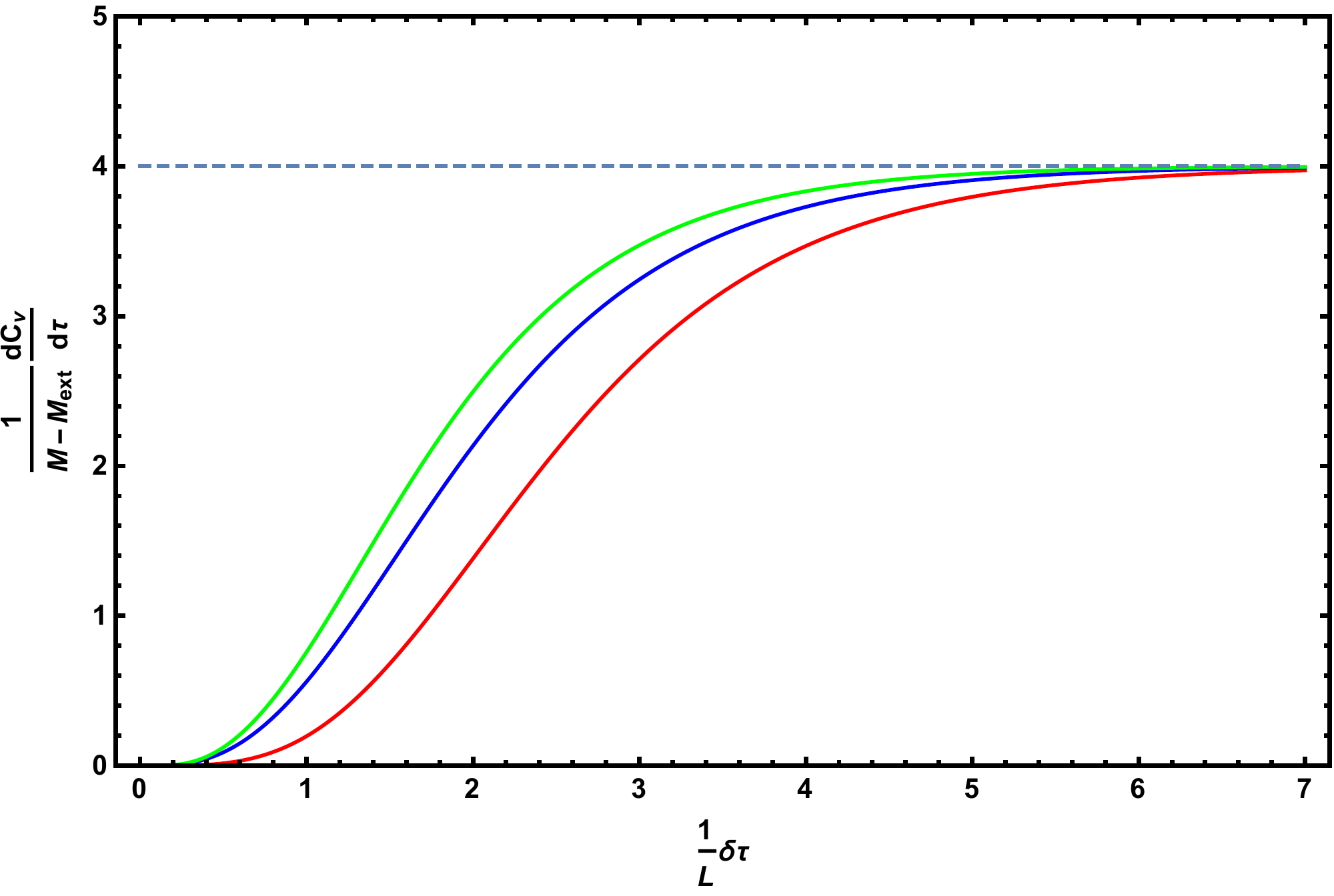}\\
		\caption{Full time evolution of complexity growth for $k=-1$. The red/blue/green curves are respectively Einstein gravity, $\tilde{\alpha}=0.04 L^{2}$, and $\tilde{\alpha}=0.08 L^{2}$. We fix $b=1$ in this case. The late time result is not affected by $\tilde{\alpha}$, but for the full time behavior, the presence of higher curvature corrections will increase the complexity growth rate which is contrary to the prediction of the $k=0$ and $k=1$ results.}	\label{ttttt}
	\end{figure}

\begin{figure}
		\includegraphics[scale=.375]{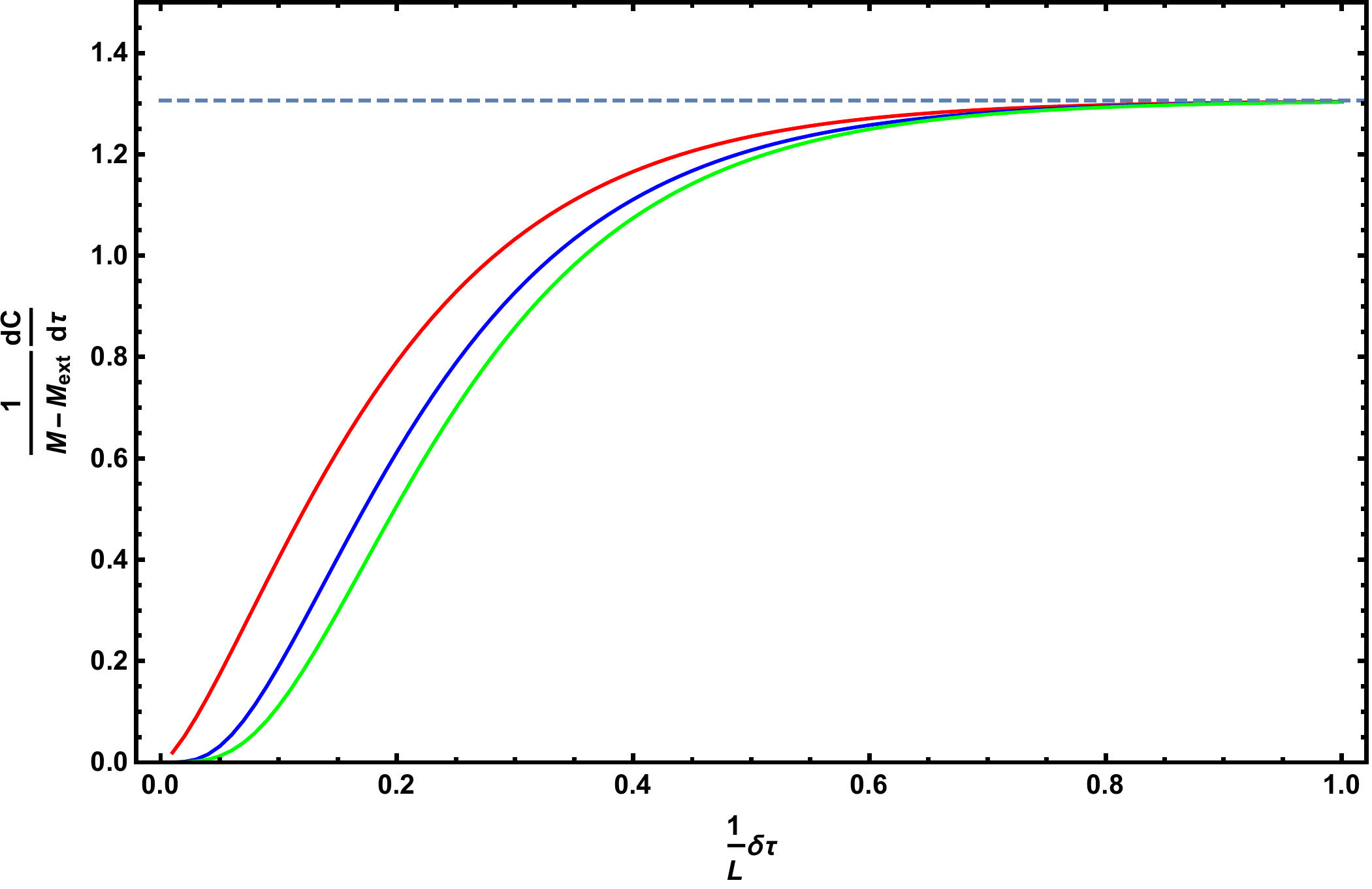}\\
		\caption{Full time evolution of complexity growth for $k=-1$. The red/blue/green curves are  Einstein gravity, $\tilde{\alpha}=0.04 L^{2}$, and $\tilde{\alpha}=0.08 L^{2}$ cases, respectively, and we fix $b=2$ here.
			The presence of the higher curvature corrections decreases the complexity growth rate for relatively large radius.}	\label{tttttferw}
	\end{figure}

	
	For $k=-1$ case, we first analyze the case when the singularity is at $ r_{s}=0$.
	The full time behavior of the $d=4, k=-1$ case with $b=1$ is plotted in Figure \ref{ttttt}, while the $d=4, k=-1$ case with $b=2$ is plotted in Figure \ref{tttttferw}. For relatively small radius, the Gauss-Bonnet coupling increases the complexity growth rate, while for relatively large radius it decreases the complexity growth rate. This is the same as the CV results.

	\subsection{The Case When $M<0$ with $k=-1$}
	As we can see, the black hole mass is always positive provided $b>1$. In this subsection, we briefly discuss the case when $M<0$, because there are two parameters that can affect the result, we can fix the radius $b$ and change the coupling $\tilde{\alpha}$. We fix $b=0.95$ in the following. 
	
\begin{figure}
		\includegraphics[scale=.375]{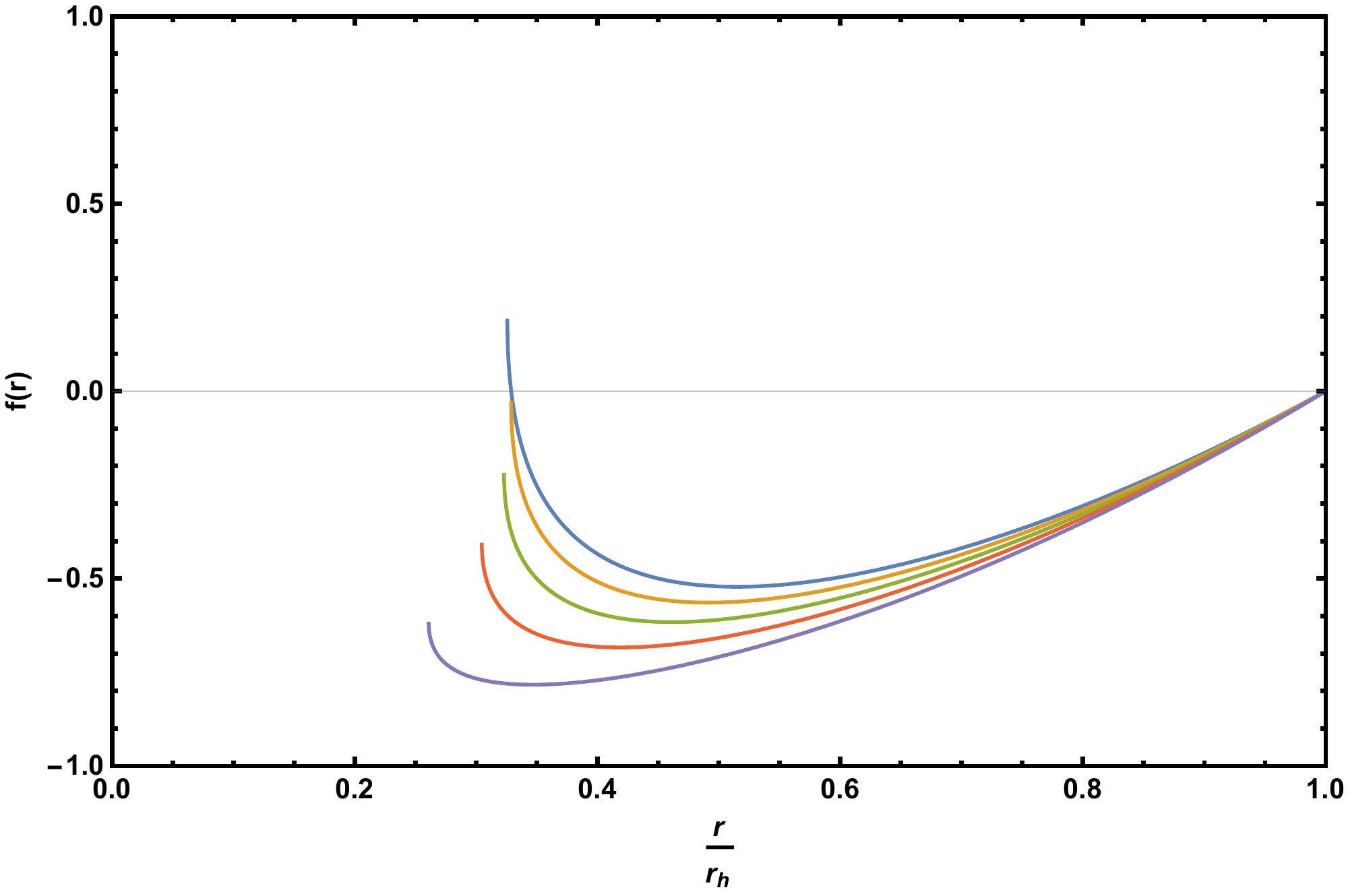}\\
		\caption{The behavior of $f(r)$ for various $\tilde{\alpha}$s, where $\tilde{\alpha}$ is respectively $0.04L^{2}$, $0.05 L^{2}$, $0.06L^{2}$, $0.07 L^{2}$, $0.08L^{2}$ from up to down.}\label{f(r)}
	\end{figure}

	In this case, we analyze the $f(r)$ for various $\tilde{\alpha}$s. It is shown in Figure \ref{f(r)}. We find that as we decrease $\tilde{\alpha}$ from its maximum $0.09 L^{2}$, there may be an inner horizon after we cross some critical value $\tilde{\alpha}_{ct}$.
	Below that critical value, the Penrose diagram looks like that of an RN-AdS black hole, and the above calculation is not correct for this case. We leave this case to further study and we want to focus on the $\tilde{\alpha}$ value which is above the critical value and see if the new singularity can lead to some new behavior of the complexity growth rate. 
	
\begin{figure}
		\includegraphics[scale=.375]{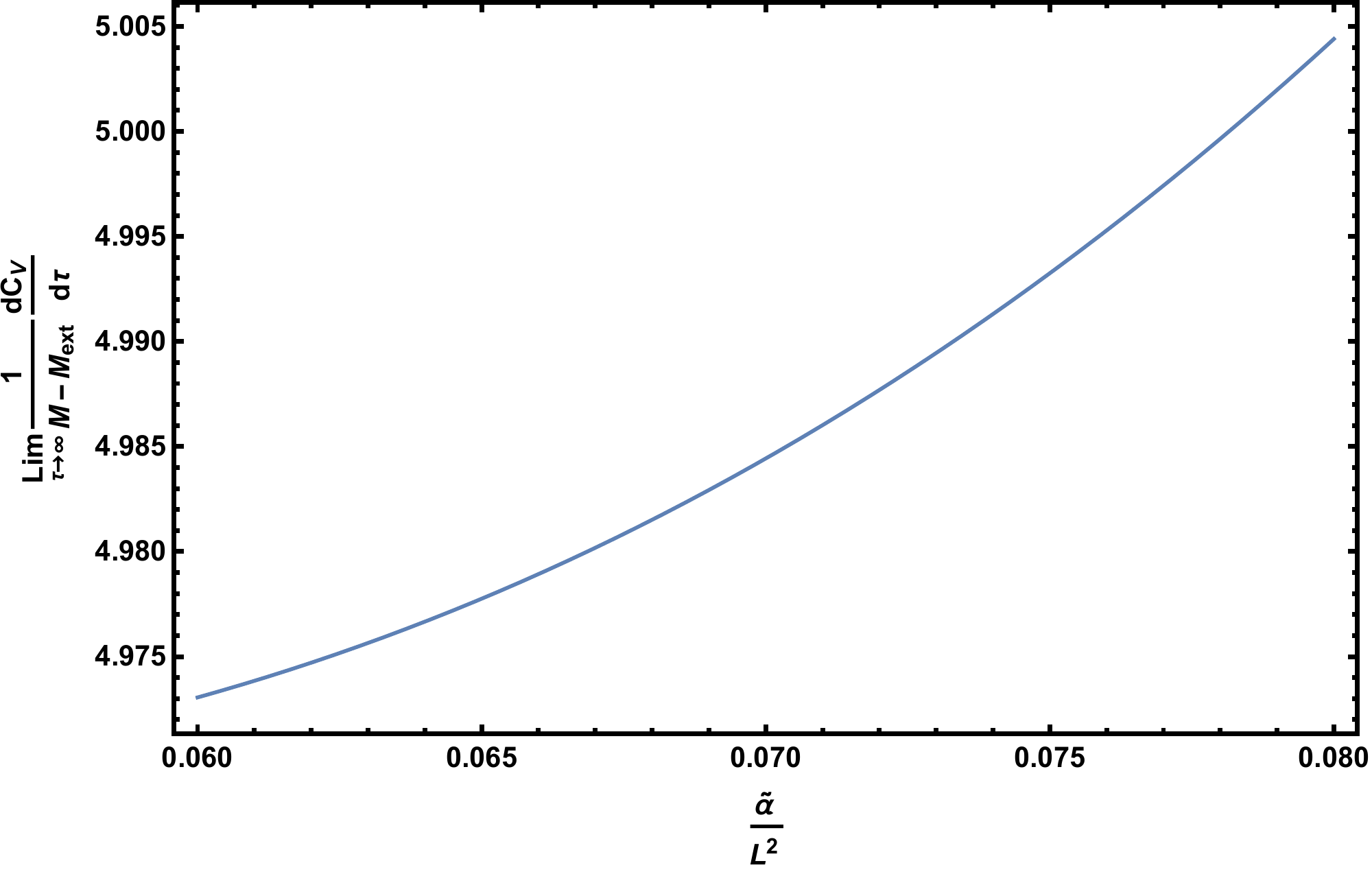}\\
		\caption{The late time complexity growth rate for $k=-1$ and $M<0$. In this case we fix $b=0.95$ and $d=4$, and we find that in this case the Gauss-Bonnet coupling indeed increases the growth rate in this range of $\tilde{\alpha}$.}\label{M<0latetime}
	\end{figure}

	We focus on the $d=4$ case, and choose $\tilde{\alpha}$ between $0.06 L^{2}$ and $0.08 L^{2}$. In this region, the Penrose diagram contains only one horizon. So the complexity growth rate is 
	\begin{equation}
		\frac{\mathrm{d}C_V}{\mathrm{d}t}=\frac{3 \Omega_{-1,3}}{16 \pi L^{2}}(r_{h}^{4}-r_{s}^{4}).
	\end{equation}
	Using the mass expression in this case
	\begin{equation}
		M=\frac{3 \Omega_{-1,3}r_{h}^{2}}{16 \pi }(-1+\frac{r_{h}^{2}}{L^{2}}+\frac{\tilde{\alpha}}{r_{h}^{2}})
	\end{equation}
	we get the analytic expression for the late time growth rate 
	\begin{equation}
		\frac{1}{M-M_{ext}}\frac{\mathrm{d}C_{V}}{\mathrm{d}\tau}=\frac{1}{\frac{1}{4}+b^{4}-b^{2}}(b^{4}-\frac{4 a b^{2}}{1-4a}(1-\frac{a}{b^{2}}-b^{2})),
	\end{equation}
	and it is plotted in Figure \ref{M<0latetime}.
	We find that the late time result is dependent on the Gauss-Bonnet coupling $\tilde{\alpha}$ and moreover the late time result increases as the coupling increases, which is similar to the CV case. 
	
	     \section{The Growth Rate of Complexity in the Generalized CV Conjectures}\label{GenVol}
	     
	     The calculations in this paper were done with the assumptions that the usual complexity volume duality holds in the Gauss-Bonnet case as in the Einstein case.
	     But at this time, because of the lack of the concrete derivation of the CV duality from field theory side, this assumption may not be true and the volume may be corrected in the presence of higher curvature corrections. 
	     In Ref.\cite{Alishahiha:2017hwg}, the authors propose two possible generalizations of the volume in CV conjecture. The first is motivated by the subregion complexity first proposed in Ref. \cite{Alishahiha:2015rta}.
	     Because the complexity is defined to be the volume between the RT surface and the boundary, as the entanglement entropy is modified in the presence of the higher curvature corrections to match the Wald entropy \cite{Wald:1993nt,Iyer:1994ys}, the author expected that the complexity should also be modified accordingly as follows
	     \begin{equation}
	     \label{cv1111}
	     C_{V1}=-\frac{1}{\ell} \int_{\Sigma} \eta E^{abcd} \epsilon_{ab} \epsilon_{cd}\,,
	     \end{equation}
	     where $\Sigma$ denotes the surface, $\eta$ is its element volume, and $E^{abcd}$ is defined as
	     \begin{align} \label{defEtensor}
	     	E^{abcd} &= \frac{\partial   L}{\partial R_{abcd}}  - \nabla_{a_1} \frac{\partial L}{\partial \nabla_{a_1} R_{abcd}} + \dots   \\
	     	&+ (-1)^m \nabla_{(a_1} \cdots \nabla_{a_m )} \frac{\partial L}{\partial  \nabla_{(a_1} \cdots \nabla_{a_m )} R_{abcd}} \nonumber  \, .
	     \end{align}
	     For a co-dimension one hypersurface, its normal vector $u_{a}$ and the vector normal to the constant radial coordinate surface $n_{a}$ form the bi-normal $\epsilon_{ab}$ in the above expression.
	     
	     The other modification appears in Ref.\cite{Bueno:2016gnv} when they investigated the ``entanglement equilibrium'' in the context of higher order gravity theories, and some generalized form of volume is kept fixed when the entanglement entropy is varied. For a co-dimension one maximal time slice, the complexity is dual to the following functionaly 
	     \begin{equation}
	     \label{cv2222}
	     C_{V2}=\frac{1}{\ell} \int_{\Sigma}\left[ E^{abcd}( a u_{a} u_{b} h_{c d}+b h_{ab} h_{cd})+ c  \right]\eta
	     \end{equation}
	     where  the coefficients $a, b$ and $c$ are constants.
	     
	     In order to know whether these volume proposals are suitable for describing complexity in Gauss-Bonnet gravity, we can calculate the complexity growth rate explicitly and try to find whether it shows the expected physical properties of complexity.
	     
	     For the Gauss-Bonnet gravity ,the Lagrangian reads
	     \begin{eqnarray}
	     L = \frac{1}{16\pi G}\left(R -2 \Lambda+  \alpha ( R^2 -  4 R_{ab} R^{ab}  +   R_{abcd}R^{abcd})\right)\,.
	     \end{eqnarray}
	     From this we have
	     \begin{equation}
	     \begin{aligned}
	     16\pi G E^{abcd} =& \, \left ( \frac{1}{2} +  \alpha R \right) 2 g^{a[c} g^{d]b} 
	     \\ & - 4\alpha \left (   R^{a[c} g^{d]b} + R^{b[d} g^{c]a} \right) + 2 \alpha R^{abcd} \, .
	     \end{aligned}
	     \end{equation}
	     To calculate the generalized volumes we need the expressions of $u^a$, $n^a$ and the Riemann curvature.
	     For the static black hole solution (\ref{staticbh}) in the Eddington coordinates, we assume that the unit normal $u^a$ has only two non-zero components
	     \begin{equation}
	     u^a = (u^v, u^r, 0,\cdots,0)
	     \end{equation}
	     and so does $n^a$.
	     These two vectors satisfy
	     \begin{eqnarray}
	     -u^a u_a=n^a n_a =1\,,\qquad u^a n_a=0\,.
	     \end{eqnarray}
	     According to the metric (\ref{Eddington}), the $vv, vr$ and $rr$ components of the Ricci tensor are
	     \begin{eqnarray}
	     R_{vv}=\frac{1}{2} f f'' + \frac{(d-1)}{2 r}f f'  R_{rr} = 0\,,
	     \end{eqnarray}
	     \begin{eqnarray}
	     R_{vr} = -\frac{1}{2}f'' - \frac{(d-1)}{2 r}f'\,,
	     \end{eqnarray}
	     \begin{eqnarray}
	     R_{rr} = 0\,,
	     \end{eqnarray}
	     where ``~$'$~'' denotes the derivative with respect to $r$,
	     and we can see that for $\alpha,\beta= v, r$, we have the simple relation
	     \begin{eqnarray}
	     R_{\alpha\beta} = -\frac{1}{2}(f''+\frac{(d-1)}{r}f') g_{\alpha\beta}\,.
	     \end{eqnarray}
	     Therefore due to the normalization $g_{ab}u^a u^b = -1$ we have
	     \begin{eqnarray}
	     2R_{ab}u^a u^b = f''+\frac{(d-1)}{r}f'\,.
	     \end{eqnarray}
	     Applying the formulae in Ref.\cite{Cai:2013cja} the Ricci scalar is
	     \begin{eqnarray}
	     R=-f''-\frac{2(d-1)}{ r} f' + (d-1)(d-2)\frac{k -f}{r^2}\,,
	     \end{eqnarray}
	     and meanwhile the $v, r$ components of the Riemann tensor of the spacetime are just the same as those of the Riemann tensor of the $v,r$-submanifold. 
	     Since this submanifold is two-dimensional, there is only one non-trivial $v,r$-component of the Riemann tensor.
	     Therefore for $\alpha,\beta, \gamma, \delta= v, r$, 
	     \begin{equation}
	     R_{\alpha\beta\gamma\delta}=- \frac{f''}{2} (g_{\alpha\gamma} g_{\beta \delta}-g_{\alpha \delta} g_{\beta \gamma})\,.
	     \end{equation}
	     Now we have all the materials to calculate the generalized volumes.
	     In order for that when $\alpha=0$ it reduces to the usual CV proposal, the generalized formula (\ref{cv1111}) turns out to be
	     \begin{equation}
	     C_{V1}= \frac{1}{G \ell} \int_{\Sigma} W_1(r)\eta
	     \end{equation}
	     where
	     \begin{eqnarray}  		\label{c333}
	     W_1(r)&=&1 + 2\alpha R +4\alpha R^{ab}\left(u_a u_b - n_a n_b\right)\nonumber\\
	     && - 4\alpha R^{abcd}u_a n_b u_c n_d \nonumber\\
	     &=&1+ 2 \alpha(d-1)(d-2) \frac{k-f}{r^{2}}
	     \,,
	     \end{eqnarray}
	     for the specific static black hole.
	     The second formula (\ref{cv2222}) gives
	     \begin{eqnarray}
	     C_{V2} =\frac{1}{G\ell} \int_{\Sigma} W_2(r)\eta
	     \end{eqnarray}
	     where
	     \begin{widetext}
	     	\begin{eqnarray}
	     	W_2(r) &=& a\left( \frac{d}{2}+(d-2)\alpha( R+2 R_{a b} u^{a} u^{b}) \right)\nonumber \\&&+b\left(\frac{d(1-d)}{2}+(d-2) \alpha((3-d)R+4R_{a b} u^{a} u^{b})\right)\nonumber\\
	     	&=& a\left(\frac{d}{2}+\alpha(d-1)(d-2)\left(-\frac{f'}{r}+(d-2)\frac{k-f}{r^2}\right)\right)\nonumber\\
	     	&&+b\left(\frac{d(1-d)}{2}+ \alpha(d-2)\left((d-1)f''+\frac{2(d-1)(d-2)f'}{r}-(d-1)(d-2)(d-3)\frac{k-f}{r^2}\right)\right)\nonumber\\
	     	&=& 1 + \frac{{\alpha  (d-2)}}{2  d} \left(2 (d-6) (d-2) \frac{k -f(r)}{r^2}-(d (d+5)-16) \frac{f'}{r}-(d+4) f''\right)
	     	\,,
	     	\end{eqnarray}
	     \end{widetext}
	     where we set $c=0$ and normalized $a$ and $b$ according to the analysis in \cite{Alishahiha:2017hwg}. The overall factor is fixed such that the expression goes back to unity when $\alpha=0$.
	     
	     For the static black hole the two generalized proposals can be written in a unified form
	     \begin{eqnarray}
	     C_{V} =\frac{ \Omega_{k,d-1}}{G\ell} \int\mathrm{d} \lambda \, r^{d-1} \sqrt{-f(r) \dot{v}^2 + 2 \dot{v}\dot{r}} W(r)\,.
	     \end{eqnarray}
	     By repeating a similar procedure as in Sec.\ref{Method} we find the growth rate of the generalized holographic complexity to be
	     \begin{eqnarray}
	     \frac{\mathrm{d}C_V}{\mathrm{d}\tau} = \frac{\Omega_{k,d-1}}{G L}\sqrt{-f(r_{\rm min})}r_{\rm min}^{d-1}\lvert W(r_{\rm min})\rvert\,.
	     \end{eqnarray}
	     
\begin{figure}
	     	     		\includegraphics[scale=.32]{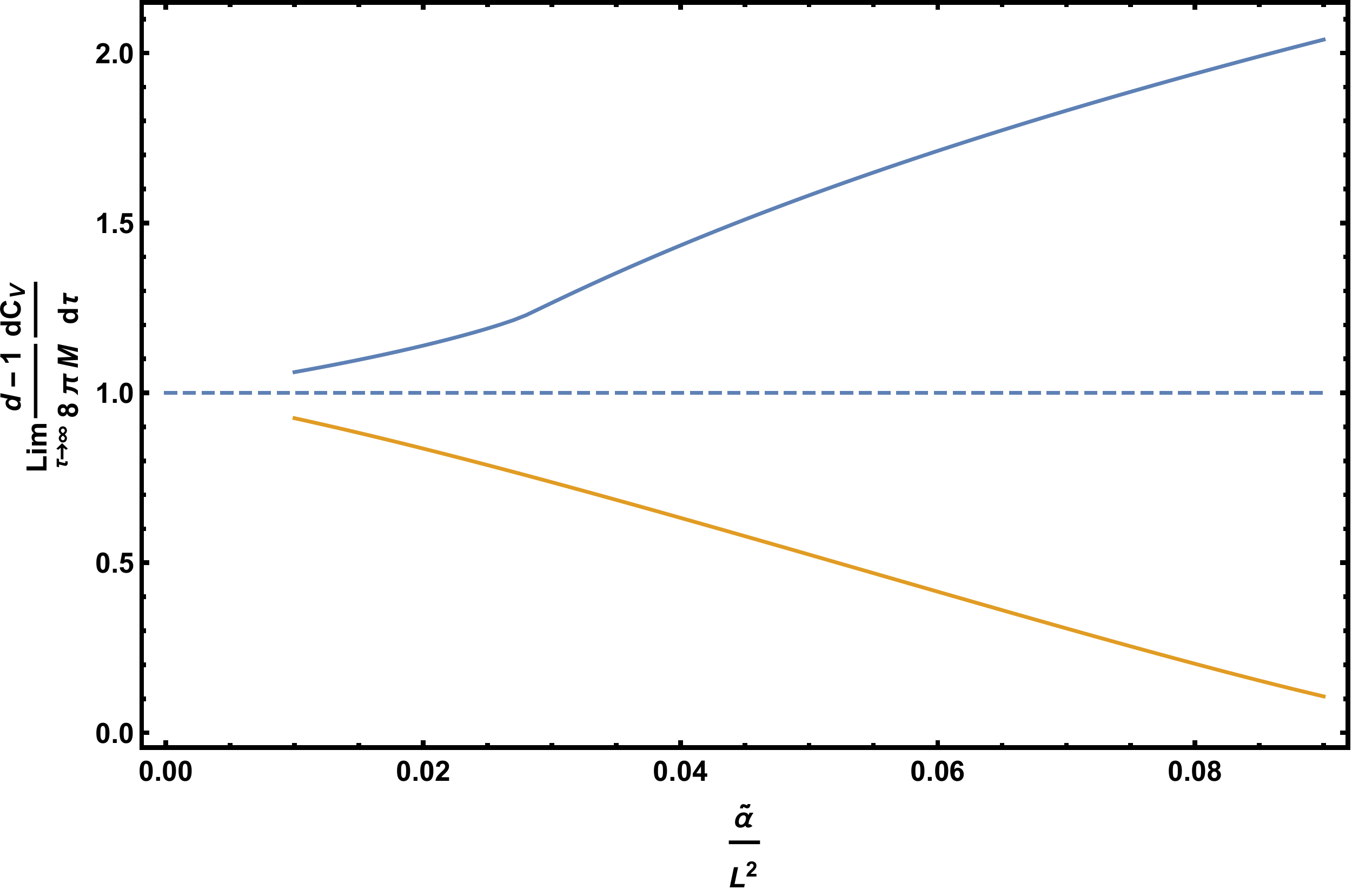}\\
	     	     		\caption{The late-time growth rate of the two generalized volume proposals for a $5$-dimensional static black hole with $k=0$. The blue curve corresponds to the first one (\ref{cv1111}) and the orange curve shows the second one (\ref{cv2222}).}	\label{Generalizedvolume}
	     	     	\end{figure}

	     {Calculations for a $5$-dimensional static black hole with $k=0$ are performed for physically allowed $\tilde{\alpha}$ range}, and the results show that for the first proposal (\ref{cv1111}), the late-time complexity growth rate exceeds that of the Einstein gravity for a positive Gauss-Bonnet parameter.
	     So this proposal is not favorable, since one should see a decrease in the complexity growth rate compared to Einstein gravity.
	     {The late-time complexity growth rate for the second proposal (\ref{cv2222}) for small Gauss-Bonnet parameters decreases drastically as the Gauss-Bonnet parameter increases. It seems that this proposal is better than the first one  (\ref{cv1111}) in this respect. It may be interesting to investigate other aspects of this proposal to see whether it is an appropriate generalization, and we leave it to the future work. }
	     These results for both proposals are shown in Figure \ref{Generalizedvolume}.
	     
{We should note that the generalized volumes are just conjectures without derivation and very strong implications, unlike the case of holographic entanglement entropy. 
It is also interesting to find other forms of generalized volumes from other physical directions.
It is also possible that we should stick to the volume of the extremal surface, since as far as the complexity growth behavior is concerned, the ``Complexity --- Volume'' conjecture shows rather good behavior. }

%

	\section{Conclusion and Discussion}\label{Discuss}
	\subsection{Summary of Our Results}

	In this work we studied the holographic complexity of AdS black holes in Einstein-Gauss-Bonnet gravity theory in the context of the ``complexity--volume'' (CV) and the CV2.0 conjectures.
	Our results include the time dependence of the complexity growth rate $\mathrm{d}C_V/\mathrm{d}\tau$ of neutral black holes with different Gauss-Bonnet couplings and different horizon curvatures ($k=0,1$ and  $k=-1$), and of charged planar black holes with different charge values.
	Our results are shown in $5$ dimensions by numerical graphs except that for the planar horizon case we obtained an analytic expression for the late time complexity growth rate in general dimensions. 
	{We also investigated two proposals of generalized volumes dual to the complexity \cite{Bueno:2016gnv,Alishahiha:2017hwg} in the presence of higher curvature corrections. We find that complexity growth rate for these two proposals behaves rather different and we find the proposal in \cite{Bueno:2016gnv} is better. However, we should note that we need further evidence for us to trust the proposal in Ref.\cite{Bueno:2016gnv}}.
	
	For all the cases we investigated, $\mathrm{d}C_V/\mathrm{d}\tau$ increases monotonically as time goes on, and approaches a certain constant from below at late times. This is also the case in Einstein gravity \cite{Carmi:2017jqz}. The monotonic growth of $\mathrm{d}C_V/\mathrm{d}\tau$ in both gravity theories implies that the ``complexity equals volume'' conjecture is more favorable than the ``complexity--action'' (CA) conjecture in the sense of the Lloyd's bound. 
	
	Our results found by using the CV conjecture also indicate that except for small Gauss-Bonnet AdS black holes with hyperbolic horizons,  the growth rate can be larger than that in Einstein gravity.
	To be more specific, we find that when $k=-1$ and $r_h/L$ is large, or when $k=0, 1$ we always have that
	\begin{eqnarray}
		\frac{d-1}{8 \pi  (M-M_{\rm ext})}\frac{ \mathrm{d}C_V}{\mathrm{d}\tau }\Big|_{\rm EGB}<\frac{d-1}{8 \pi  (M-M_{\rm ext})}\frac{ \mathrm{d}C_V}{\mathrm{d}\tau }\Big|_{\rm Einstein}\,,\nonumber\\
	\end{eqnarray}
	which is expected. 
	However when $k=-1$ and $r_h/L$ is small, one may have
	\begin{eqnarray}
		\frac{d-1}{8 \pi  (M-M_{\rm ext})}\frac{ \mathrm{d}C_V}{\mathrm{d}\tau }\Big|_{\rm EGB}>\frac{d-1}{8 \pi  (M-M_{\rm ext})}\frac{ \mathrm{d}C_V}{\mathrm{d}\tau }\Big|_{\rm Einstein}\,.\nonumber\\
	\end{eqnarray}
	
	As for the CV2.0 proposal, for the $k=0$ case, while the late time result is the same as in the Einstein gravity, the higher curvature corrections decrease the growth rate when the time is not too late. For the $k=1$ case, the higher order corrections decrease the growth rate both at late times and in full time. In the $k=-1$ case, we should distinguish the case between the $M>0$ case and $M<0$ case.
	For the $M>0$ case, we find that the higher curvature corrections decrease the complexity growth rate for relatively large black holes, while increase the growth rate for relatively small black holes. And the late time limit is independent of $\tilde{\alpha}$.
	However, in the $M<0$ case, the complexity growth rate is enhanced even in the late time limit.
	
	According to the AdS/CFT correspondence, higher curvature terms in the bulk correspond to the large N or large coupling constant corrections in the dual boundary field theory. 
	For example, in the AdS$_5$/CFT$_4$ case of  the type IIB string theory,  the terms with $R^4$  order in the bulk give the  correction with the 't Hooft coupling $\lambda ^{-3/2}$ in 
	the boundary field theory~\cite{Gubser:1998nz} . It is expected that with those corrections in the bulk and boundary, one is able to make a comparison of calculations from the bulk 
	and the boundary. 
	 In this work we have considered the effect of the Gauss-Bonnet term in the bulk.  On one hand, the Gauss-Bonnet gravity is a natural extension of Einstein gravity 
	in high dimensional spacetime,  one can have an analytical black hole solution in the Gauss-Bonnet gravity and the vacuum of the theory 
	is stable and the theory has no ghost ~\cite{Boulware:1985wk,Cai:2001dz} .   On the other hand,  the Gauss-Bonnet term is a low energy correction term in the heterotic string 
	theory.  Therefore considering the effect of the Gauss-Bonnet term is of some interest not only in its own right in the sense of gravity theory itself, but also in checking the AdS/CFT
	correspondence and/or in understanding the properties of strong coupling field theory with the AdS/CFT correspondence. 
	Our study in this paper shows that for the cases with 
	$k=0$ and $k=1$, the Gauss-Bonnet term always suppresses the complexity growth rate for both the late time limit and the full time evolution cases. This conclusion agrees with 
	the one from the CA conjecture~\cite{Cai:2016xho}, and it is also expected from the field theory side  considering the Gauss-Bonnet term as some correction of large N 
	expansion~\cite{Gubser:1998nz}.  In particular, it is speculated that stringy corrections should reduce the complexity growth rate of the AdS black hole solutions~\cite{Brown:2015lvg}.
	
    On the other hand, the enhancement we find in the $k=-1$ case is opposite to what is expected.
    This unexpected behavior makes us recall the fact  that the boundary field theory in a hyperbolic space is not well-behaved, as argued in \cite{Witten:1999xp}. Therefore this unexpected behavior might be not trustable. 
	
	We note that it is still necessary to investigate the complexity growth rate in the very weak coupling case from field theory side to complete the whole analysis.

	
	\subsection{Comparison to the CA Result}
	
	It would be helpful to compare our results with those obtained by using the CA conjecture. The growth rate in the context of CA conjecture for a spherical black hole in EGB gravity with $d=4$ is \cite{Cai:2016xho}
	\begin{eqnarray}\label{GBCA}
		\lim\limits_{\tau\rightarrow \infty}\frac{\pi}{2M}\frac{\mathrm{d} C_A}{\mathrm{d}\tau} = \left(1 - \frac{3\tilde{\alpha}\Omega_{1,4}}{16\pi G M}\right)\,.
	\end{eqnarray}
	As our results, this growth rate also contains a suppression compared to that in Einstein gravity. Recent studies on the null boundary action term in Lovelock gravity \cite{Cano:2018ckq,Cano:2018aqi} give a more general result. Nevertheless, the suppression appearing in this expression is different from our result. This suppression only appears when $d$ is even and $k\neq 0$ \cite{Cano:2018aqi}. In our CV results, the suppression appears for all three values of $k$, and it appears for any $d$. Besides, we do not have such an analytic expression as (\ref{GBCA}) for $k= 1$ case in the CV conjecture.
	Moreover, under the large black hole limit, the correction term $- {3\tilde{\alpha}\Omega_{1,4}}/({16\pi G M})$ vanishes, which means that the effect of the corrections will disappear for large black holes. In our late time results in the CV conjecture, however, the effect of $\tilde{\alpha}$ becomes the same as in the case of $k=0$ under the limit $r_h/L\rightarrow\infty$ and this effect is always finite. 
	The suppression from the Gauss-Bonnet coupling appears in both cases while there are curious differences between these cases --- the suppression seems to be more universal in the CV conjecture.
	Such investigation may help judge which holographic proposal captures the essential features of complexity.
	We can give a summary of the result in Table \ref{my-label}.
	
	\begin{table}[]
		\centering
		\caption{Effect of higher curvature corrections in various proposals}
		\label{my-label}
		\begin{tabular}{|l|l|l|l|l|}
			\hline
			&      & CV                                                                                                  & CV2.0                                                                                                                  & CA                                                             \\ \hline
			\multirow{3}{*}{\begin{tabular}[c]{@{}l@{}}late \\ \\ time\end{tabular}}   & k=0  & decrease                                                                                            & unchanged                                                                                                              & unchanged                                                      \\ \cline{2-5} 
			& k=1  & decrease                                                                                            & decrease                                                                                                               & \begin{tabular}[c]{@{}l@{}}decrease for \\ even d,\\unchanged for\\ odd d\end{tabular} \\ \cline{2-5} 
			& k=-1 & \begin{tabular}[c]{@{}l@{}}decrease for large\\  radius\\ increase for small \\ radius\end{tabular} & \begin{tabular}[c]{@{}l@{}}unchanged \\ for M\textgreater{}0 \,(d=4)\end{tabular}                                              & \begin{tabular}[c]{@{}l@{}}decrease for \\ even d,\\unchanged for\\ odd d\end{tabular} \\ \hline
			\multirow{3}{*}{\begin{tabular}[c]{@{}l@{}}full\\ \\ \\ time\end{tabular}} & k=0  & decrease                                                                                            & decrease                                                                                                               & \multirow{3}{*}{Unknown}                                       \\ \cline{2-4}
			& k=1  & decrease                                                                                            & decrease                                                                                                               &                                                                \\ \cline{2-4}
			& k=-1 & \begin{tabular}[c]{@{}l@{}}decrease for large \\ radius\\ increase for small \\ radius\end{tabular} & \begin{tabular}[c]{@{}l@{}}decrease for large \\ radius\\ increase for small\\ radius  (M\textgreater{}0)\end{tabular} &                                                                \\ \hline
		\end{tabular}
	\end{table}

	\subsection{Choice of Boundary Time Coordinate and Future Directions}
	\label{lastdiscuss}

In the above discussion, we have not considered what the boundary time should be in the presence of the Gauss-Bonnet coupling.
Take the black brane as an example. We note that under the limit $r \to \infty$, in order to set the speed of light on the boundary equal to $1$ \cite{Brigante:2007nu,Roberts:2014isa}, we should shift the time coordinate $t\rightarrow t'$ according to 
$t=Nt'$
where 
\begin{equation}
	N=\sqrt{\frac{1}{2}(1+\sqrt{1-4\tilde{\alpha}/{L^{2}}})} \,.
\end{equation}
Physically, we should use $t'$ to be the boundary time and compute the boundary complexity growth rate
\begin{equation}
	\frac{{\rm d}C}{{\rm d}t'}=N \frac{{\rm d}C}{{\rm d}t}\,.
\end{equation}
Since $N<1$ so the complexity growth rate should decrease more. 
This decrease is naturally expected because of the following reason.

In Ref.\cite{Roberts:2014isa}, the authors investigated the localized shocks and the Gauss-Bonnet coupling's effect on the butterfly velocity. In fact, that paper showed that for two space separated perturbations $V_{x}(t)$ and $W_{y}(0)$, the commutator between them takes the form
\begin{equation}
	-\langle [V_{x}(t),W_{y}(0) ]^{2} \rangle= \frac{1}{N^{2}} e^{\frac{2 \pi}{\beta}(t-  |x-y|/v_B)}
\end{equation}
which gives a natural light cone of scrambling in terms of the butterfly velocity $v_B$ satisfying
\begin{equation}
	v_{B}=\frac{2\pi}{\beta \mu}\,,\qquad \mu=\sqrt{d(d-1)/2}\,.
\end{equation}
Although the scrambling time and butterfly velocity take the same form in the Gauss-Bonnet case as in Einstein gravity, a constant scaling of the time coordinates which set the speed of light on the boundary changes the value of $\beta$.
So the butterfly velocity is decreased in the presence of the Gauss-Bonnet coupling to the value
\begin{equation}
	v_{B}=\frac{1}{2} \sqrt{1+\sqrt{1-4\tilde{\alpha}/L^{2}}} \sqrt{\frac{d}{d-1}}
\end{equation}
while the result in Einstein gravity is $v_{B}=\sqrt{{d}/{(2(d-1))}}$.
This is the same behavior as we found for the complexity growth, and is natural because complexity growth and chaos is closely related.
More concretely, using the tensor network picture in \cite{Brown:2015lvg}, smaller $v_{B}$ means smaller rate of growth of complexity of the precursor operator.

Finally let us talk about future directions.
Firstly, we have only studied the effect of the Gauss-Bonnet term to the complexity growth rate, since the single parameter makes the problem clear and simple. It might be straightforward to do the same calculations in more general higher curvature or higher derivative gravity theories, for example the Lovelock gravity theory.

Secondly, it is tempting to compare the time dependence of the complexity in both the CV and CA methods, as the papers \cite{Carmi:2017jqz,Kim:2017qrq} did, in the presence of higher curvature or higher derivative terms.
However we are still not able to answer what the full time behavior of the complexity will be for the higher curvature gravity theories in the context of the CA conjecture, since the contribution of the action on the null boundaries of the Wheeler-DeWitt patch is yet unknown, although some progress has been made to identify the contribution of the joints connecting two boundary sections of the Wheeler-DeWitt patch in the paper \cite{Cano:2018ckq} for Lovelock gravity. The first step is to obtain proper boundary action for these theories, which is an important and challenging work.

Moreover, how to interpret the effect of higher curvature corrections on the complexity growth is still interesting. An initial attempt of understanding the relation between the butterfly velocity and complexity had been made in Ref.\cite{HosseiniMansoori:2017tsm}.
But as we showed in the main body of this paper, even we do not take into account the effect of the slowing down of scrambling in the black brane case (the effect of rescaling the boundary time), the complexity growth rate is still decreased.
So the complexity growth and scrambling may not slow down due to the same reason.

	\begin{acknowledgments}
		We thank Li-Ming Cao and Run-Qiu Yang for valuable suggestions and discussions.
		This work is supported in part by the National Natural Science Foundation of China Grants No.11690022, No. 11435006, No.11447601 and No.11647601, and by the Strategic Priority Research Program of CAS Grant No.XDB23030100, and by the Peng Huanwu Innovation Research Center for Theoretical Physics Grant No.11747601, and by the Key Research Program of Frontier Sciences of CAS. 
		Yuxuan Peng is supported in part by the National Postdoctoral Program for Innovative Talents Grant No. Y7Y2351B11.
	\end{acknowledgments}

\end{document}